\documentclass[aps,prb,twocolumn,amsmath,amssymb,superscriptaddress]{revtex4-2}

\usepackage{graphicx}

\usepackage{xcolor}

\usepackage{siunitx}

\begin{document}

\title{Transverse magnetic focusing in two-dimensional hole gases}

\affiliation{Theoretical, Computational, and Quantum Physics, School of Science, RMIT University, Melbourne, Australia}
\affiliation{School of Physics, University of New South Wales, Sydney, NSW 2052, Australia}
\affiliation{ARC Centre of Excellence in Future Low-Energy Electronics Technologies, Australia}

\author{Yik K. Lee}
\email{yik.kheng.lee@rmit.edu.au}
\affiliation{Theoretical, Computational, and Quantum Physics, School of Science, RMIT University, Melbourne, Australia}
\affiliation{ARC Centre of Excellence in Future Low-Energy Electronics Technologies, Australia}
\author{Jackson S. Smith}
\affiliation{Theoretical, Computational, and Quantum Physics, School of Science, RMIT University, Melbourne, Australia}
\affiliation{ARC Centre of Excellence in Future Low-Energy Electronics Technologies, Australia}
\author{Hong Liu}
\affiliation{School of Physics, University of New South Wales, Sydney, NSW 2052, Australia}
\affiliation{ARC Centre of Excellence in Future Low-Energy Electronics Technologies, Australia}
\author{Dimitrie Culcer}
\affiliation{School of Physics, University of New South Wales, Sydney, NSW 2052, Australia}
\affiliation{ARC Centre of Excellence in Future Low-Energy Electronics Technologies, Australia}
\author{Oleg P. Sushkov}
\affiliation{School of Physics, University of New South Wales, Sydney, NSW 2052, Australia}
\affiliation{ARC Centre of Excellence in Future Low-Energy Electronics Technologies, Australia}
\author{Alexander R. Hamilton}
\affiliation{School of Physics, University of New South Wales, Sydney, NSW 2052, Australia}
\affiliation{ARC Centre of Excellence in Future Low-Energy Electronics Technologies, Australia}
\author{Jared H. Cole}
\email{jared.cole@rmit.edu.au}
\affiliation{Theoretical, Computational, and Quantum Physics, School of Science, RMIT University, Melbourne, Australia}
\affiliation{ARC Centre of Excellence in Future Low-Energy Electronics Technologies, Australia}

\begin{abstract} 
Two-dimensional hole gases (2DHGs) have strong intrinsic spin-orbit coupling and could be used to build spin filters by utilising transverse magnetic focusing (TMF)~\cite{Heremans1992, Heremans1994, Rokhinson2004, Rendell2015, Rendell2023}.
However, with an increase in the spin degree of freedom, holes demonstrate significantly different behaviour to electrons in TMF experiments, making it difficult to interpret the results of these experiments. 
In this paper, we numerically model TMF in a 2DHG within a GaAs/Al\textsubscript{x}Ga\textsubscript{1-x}As heterostructure.
Our band structure calculations show that the heavy $(\langle J_{z} \rangle = \pm\frac{3}{2})$ and light $(\langle J_{z} \rangle = \pm\frac{1}{2})$ hole states in the valence band mix at finite $k$, and the heavy hole subbands which are spin-split due to the Rashba effect are not spin-polarised.
This lack of spin polarisation casts doubt on the viability of spin filtering using TMF in 2DHGs within conventional GaAs/Al\textsubscript{x}Ga\textsubscript{1-x}As heterostructures.
We then calculate transport properties of the 2DHG with spin projection and offer a new perspective on interpreting and designing TMF experiments in 2DHGs.
\end{abstract}

\maketitle

\section{Introduction}
\label{section:Introduction}
    When electrons are injected into a two-dimensional electron gas (2DEG) in the presence of a perpendicular magnetic field ($B^{}_{z}$), they follow cyclotron orbits~\cite{Tsoi1999}.
    If Rashba spin-orbit interaction (SOI) is present, the electrons follow spin dependent paths, resulting in spin polarised conductance.
    This transverse magnetic focusing (TMF) technique is well-studied in the literature and has been shown to be a viable spin filter in 2DEGs~\cite{Potok2002,Folk2003,Rokhinson2004,Dedigama2006,Heremans2007,Bozhko2014,Lo2017,Yan2017,Yan2018}.
    
\begin{figure}[htp]
        \centering
        \includegraphics[width=0.9\linewidth]{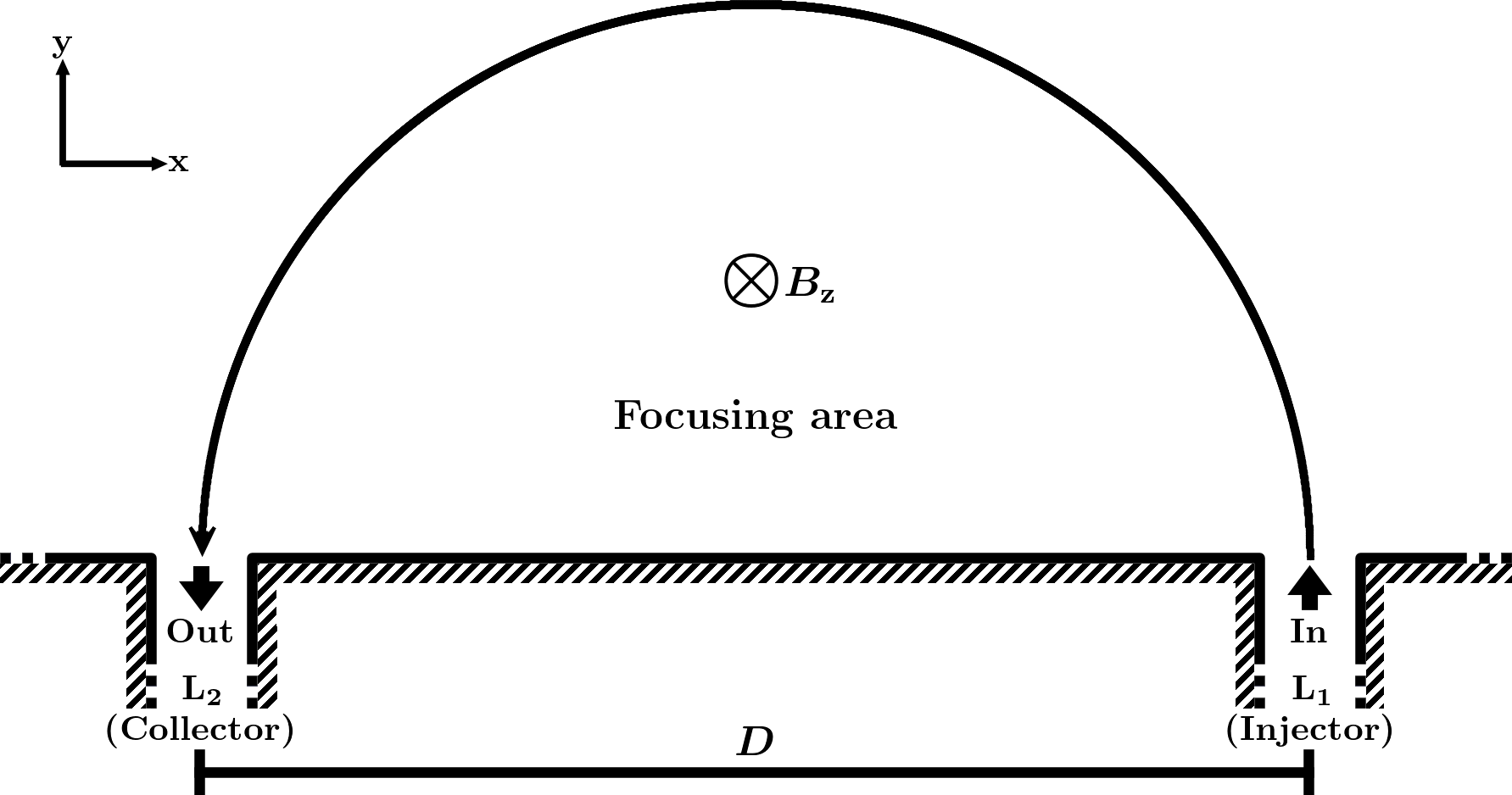}   
        \caption{
        Schematic of a typical transverse magnetic focusing experimental setup.
        Carriers are injected through an injector lead ($\mathrm{L}_1$) into a scattering area where a perpendicular magnetic field $B_z$ is applied.
        Tuning the magnetic field allows the carriers to be focused onto the collector lead ($\mathrm{L}_2$) when the diameter of the cyclotron orbit coincides with the distance between the injector and collector leads ($D$).
        }
        \label{fig:TMF_system}
    \end{figure}

    A typical TMF experiment is setup as shown in Fig.~\ref{fig:TMF_system}, where the resistance or conductance between the injector and collector leads is measured as the magnetic field strength is varied.
    This results in a conductance spectrum with peaks at magnetic field strengths where the diameter of the cyclotron orbit coincides with the distance between the injector and collector leads ($D$).
    These peaks in the spectrum occur at~\cite{vanHouten1989}
    \begin{equation}
        B^{}_{z}=\frac{2\hbar k}{eD}
        \label{Eq:larmor}
    \end{equation}
    where $\hbar$ is the reduced Planck constant, $k$ is the magnitude of the electron's wavevector, and $e$ is the elementary charge. 
    
    When Rashba SOI couples spin and momentum, we see that electrons of different spins focus at the collector at different $B^{}_{z}$.
    This is the origin of the spin filter effect.
    However, 2DEGs formed in commonly used n-type heterostructures usually have weak spin-orbit coupling, thus making it difficult to resolve the spin-separated peaks in the conductance spectra.
    An alternative is to use p-type heterostructures which form two-dimensional hole gases (2DHGs) that have strong intrinsic SOI~\cite{Heremans1992, Heremans1994, Rokhinson2004, Rendell2015, Rendell2023}.
    Due to the complexity of the valence band in these materials, TMF experiments in 2DHGs show significantly different behaviour to 2DEGs. 
    Interpreting these results is an ongoing challenge as simple ideas using a single band effective mass with Rashba SOI are insufficient, as we will demonstrate. 
    
    In this paper, we start from a $4\times4$ Luttinger Hamiltonian~\cite{Luttinger1955} describing a valence band consisting of heavy $(\langle J_{z} \rangle = \pm\frac{3}{2})$ and light $(\langle J_{z} \rangle = \pm\frac{1}{2})$ holes, and perform numerical simulations of TMF in a 2DHG in a GaAs/Al\textsubscript{x}Ga\textsubscript{1-x}As heterostructure.
    First we calculate the band structure of a 2DHG with spin projection in Sec.~\ref{section:bandstructure}, and show that mixing occurs between the heavy and light hole states even at small $k$ values.
    Thus, while the degeneracy of the heavy hole subbands is lifted due to the Rashba effect, those subbands are not spin-polarised.
    Therefore, the common interpretation that the peaks in the conductance spectra measured in 2DHG TMF experiments are spin-polarised is not accurate.
    In Sec.~\ref{section:hardwall}, we model a 2DHG TMF device to calculate conductance spectra with spin projection and show that the peaks are indeed not spin-polarised.
    We explore the features observed in the conductance spectra by modifying our model to include quantum point contacts (QPCs) in Sec.~\ref{section:QPC}, applying disorder in Sec.~\ref{section:Disorder}, and turning off the Rashba effect in Sec.~\ref{section:Square well}.
    In doing so we provide fresh insight into the interpretation and design of TMF experiments in 2DHGs.

\section{Computational model of transport in 2DHG devices}
\label{section:Methods}
    In this section, we develop an effective Hamiltonian for a 2DHG as a function of the confinement potential at the interface of the  heterostructure. 
    We then use this effective Hamiltonian to compute conductance spectra for a range of experimentally relevant configurations.
    Throughout this paper we utilise the python package Kwant~\cite{Groth2014} to perform calculations of the transmission and LDOS (see section~\ref{subsection:computational}).
    
    \subsection{Hamiltonian}
        \subsubsection{Luttinger Hamiltonian}
            In 1954, Luttinger and Kohn developed an effective mass Hamiltonian that describes six valence bands and two conduction bands in semiconductors~\cite{Luttinger1955}.
            We are interested in the regime where the majority of carriers are heavy holes (HH), and interactions between heavy holes and light holes (LH) are possible.
            Therefore, we choose to model our 2DHG systems with a 4{\texttimes}4 Luttinger Hamiltonian ($H^{}_{\mathrm{L}}$) that only describes the relevant valence bands~\cite{Fishman1995}.
            
            We express $H^{}_{\mathrm{L}}$ in the basis of $J^{}_{z}$ eigenstates $\left\{ |{+\frac{3}{2}}\rangle, |{-\frac{3}{2}}\rangle, |{+\frac{1}{2}}\rangle, |{-\frac{1}{2}}\rangle \right\}$, following the basis used in the supplementary material for Liu~et~al., 2018~\cite{Liu2018}.     
            First, the following terms are defined:
            \begin{align*}
                P &= \frac{\mu}{2}\gamma^{}_{1}\left( k^{2}_{} + k^{2}_{z} \right)\\
                Q &= -\frac{\mu}{2}\gamma^{}_{2}\left( 2k^{2}_{z} - k^{2}_{} \right)\\
                L &= -\sqrt{3}\mu\gamma^{}_{3}k^{}_{-}k^{}_{z}\\
                M &= -\frac{\sqrt{3}\mu}{2}\left[\gamma^{}_{2}\left(k^{2}_{x} - k^{2}_{y}\right) - 2i\gamma^{}_{3}k^{}_{x}k^{}_{y} \right] 
            \end{align*}
            where ${\mu=-\frac{\hbar^{2}_{}}{m^{}_{0}}}$, $m^{}_{0}$ is the electron rest mass, ${k^{2}_{}=\sqrt{k^{2}_{x}+k^{2}_{y}}}$, ${k^{}_{\pm}={k^{}_{x} \pm ik^{}_{y}}}$, and $\gamma^{}_{1}$, $\gamma^{}_{2}$, and $\gamma^{}_{3}$ are the Luttinger parameters.

            \begin{figure}[htp]
                \centering
                \includegraphics[width=0.95\linewidth]{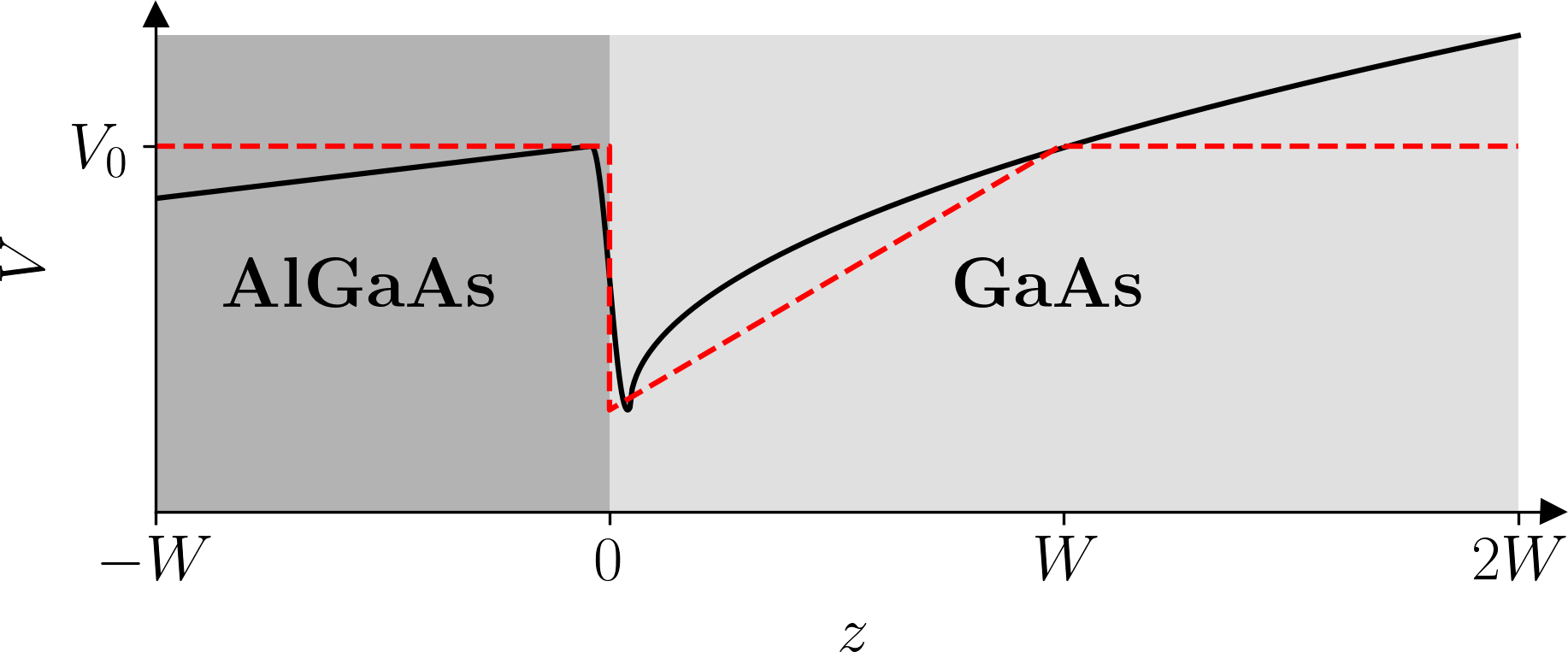}
                \caption{
                Diagram of the confinement potential at the interface of a GaAs/Al\textsubscript{x}Ga\textsubscript{1-x}As heterostructure (black solid line).
                The red dashed line shows a triangular potential well used in our calculations to approximate the confinement potential.
                }      
                \label{fig:2DHG_potentialwells}
            \end{figure}
            
            Then $H^{}_{\mathrm{L}}$ can be written:
            \begin{align}
                \label{Eq:2DHG_LuttingerH}
                &\hspace{4.7em}\begin{matrix}
                    |{+\frac{3}{2}}\rangle & \hspace{8pt}|{-\frac{3}{2}}\rangle & \hspace{8pt}|{+\frac{1}{2}}\rangle & \hspace{8pt}|{-\frac{1}{2}}\rangle
                \end{matrix}\nonumber\\
                H^{}_{\mathrm{L}} &=
                    \begin{matrix}
                        \langle{+\frac{3}{2}}| \\ \langle{-\frac{3}{2}}| \\ \langle{+\frac{1}{2}}| \\ \langle{-\frac{1}{2}}|
                    \end{matrix} 
                    \begin{bmatrix}
                            P+Q & 0 & L & M\\
                            0 & P+Q & M^{*}_{} & -L^{*}_{}\\
                            L^{*}_{} & M & P-Q & 0\\
                            M^{*}_{} & -L & 0 & P-Q
                    \end{bmatrix}
            \end{align}

        \subsubsection{Removing $k^{}_{z}$ dependence}
        \label{subsubsection:removing k_z}
            In semiconductor heterostructure devices, 2DHGs are formed at the interface between different materials in the heterostructure by the confining potential at the interface, which can be modified via doping or applied potentials.
            In this paper, we consider a 2DHG formed in a GaAs/Al\textsubscript{x}Ga\textsubscript{1-x}As heterostructure as shown in Fig.~\ref{fig:2DHG_potentialwells}.
            We approximate the confinement potential as a finite triangular well (Fig.~\ref{fig:2DHG_potentialwells}) described by
            \begin{align}
                \label{Eq:2DHG_Vz_numerical}
                V(z) =
                \begin{cases}
                    \infty & \text{for} \quad z < -W\\
                    V^{}_{0} & \text{for} \quad -W < z < 0\\
                    \frac{V^{}_{0}}{W}z & \text{for} \quad 0 < z < W\\
                    V^{}_{0} & \text{for} \quad W < z < 2W\\
                    \infty & \text{for} \quad z > 2W
                \end{cases}
            \end{align}
            where we set $V^{}_{0}$ to 211~meV.
            This value of $V^{}_{0}$ is half of the difference between the band gaps in $\text{Al}^{}_{0.33}\text{Ga}^{}_{0.67}\text{As}$ and GaAs, which are 1.845~eV~\cite{AlGaAs} and 1.424~eV~\cite{GaAs} respectively.
            $W$ was set to 100~nm so that the potential gradient approximated the band edge of a GaAs/$\text{Al}^{}_{0.33}\text{Ga}^{}_{0.67}\text{As}$ heterojunction~\cite{Rendell2023}.
            
            We assume that the 2DHG is sufficiently confined in $z$ such that the holes are in the ground state of the potential well in the $z$-direction.
            The bulk Luttinger Hamiltonian can then be transformed to an effective 2D Hamiltonian by reducing the $k_z$ operators in Eq.~\ref{Eq:2DHG_LuttingerH} to scalar factors through integration.
            Using $m^{}_{0}/\left(\gamma^{}_{1} - 2\gamma^{}_{2}\right)$ and ${m^{}_{0}}/\left(\gamma^{}_{1} + 2\gamma^{}_{2}\right)$ as the effective masses of the HH and LH respectively, we numerically solve the time-independent Schr\"odinger equation~(TISE) with Eq.~\ref{Eq:2DHG_Vz_numerical} as the potential term to obtain wave vectors corresponding to the heavy ($\phi^{}_{\mathrm{h}}$) and light ($\phi^{}_{\mathrm{l}}$) holes.
            The resulting ground state eigenvectors can then be used to find expectation values of $k^{}_{z}$ and $k^{2}_{z}$ in Eq.~\ref{Eq:2DHG_LuttingerH}, allowing us to rewrite Eq.~\ref{Eq:2DHG_LuttingerH} as~\cite{Liu2018}
            \begin{align}
                \label{Eq:2DHG_Hamiltonian_2D}
                H^{}_{2\mathrm{D}}(x,y) &=
                \begin{bmatrix}
                H^{}_{11} & 0 & H^{}_{13} & H^{}_{14}\\
                0 & H^{}_{11} & H^{}_{23} & H^{}_{24}\\
                H^{\dagger}_{13} & H^{\dagger}_{23} & H^{}_{33} & 0\\
                H^{\dagger}_{14} & H^{\dagger}_{24} & 0 & H^{}_{33}
                \end{bmatrix}
            \end{align}
            where
            \begin{align*}
                H^{}_{11} &= \frac{\mu\left(\gamma_1-2\gamma_2\right)}{2}\langle\phi_{\text{h}}|k^2_z|\phi_{\text{h}}\rangle + \frac{\mu\left(\gamma_1+\gamma_2\right)}{2}k^2 \\
                H^{}_{33} &= \frac{\mu\left(\gamma_1+2\gamma_2\right)}{2}\langle\phi_{\text{l}}|k^2_z|\phi_{\text{l}}\rangle + \frac{\mu\left(\gamma_1-\gamma_2\right)}{2}k^2 \\
                H^{}_{13} &= -\sqrt{3}\mu\gamma_3k_-\langle\phi_{\text{h}}|k_z|\phi_{\text{l}}\rangle\\
                H^{}_{14} &= M\langle\phi_{\text{h}}|\phi_{\text{l}}\rangle\\
                H^{}_{23} &= M^{*}\langle\phi_{\text{h}}|\phi_{\text{l}}\rangle\\
                H^{}_{24} &= \sqrt{3}\mu\gamma_3k_+\langle\phi_{\text{h}}|k_z|\phi_{\text{l}}\rangle
            \end{align*}
            where the Luttinger parameters are those for bulk GaAs~\cite{Winkler2003} $(\gamma^{}_{1}=6.85, \gamma^{}_{2}=2.10$, and $\gamma^{}_{3}=2.90)$.

            We should note here that the terms heavy holes and light holes are ill-defined in this effective Hamiltonian.
            In the definitions of $H^{}_{11}$ and $H^{}_{33}$ in Eq.~\ref{Eq:2DHG_Hamiltonian_2D}, ``heavy'' and ``light'' in $\phi^{}_{\mathrm{h}}$ and $\phi^{}_{\mathrm{l}}$ refers to the effective masses associated with the $k^{}_{z}$ terms, which are inversely proportional to $\gamma_1 - 2\gamma_2$ and $\gamma_1 + 2\gamma_2$ respectively.
            The commonly used definition of heavy and light holes in literature~\cite{Winkler2003} correspond to the $\pm\frac{3}{2}$ (heavy) and $\pm\frac{1}{2}$ (light) states in the $J^{}_{z}$ basis.

            However, when we consider transport in the 2D plane, it is the $k^{2}$ terms in $H^{}_{11}$ and $H^{}_{33}$ which are relevant, and so the effective masses to be considered are inversely proportional to $\gamma_1 \pm \gamma_2$.
            So in the 2D plane, the heavy hole as previously defined would have a smaller effective mass than a light hole.
            It is therefore important to keep this difference in mind when computing quantities like the cyclotron radius or mean free path.
            In addition, in the 2D plane, the simple assignment of $\pm\frac{3}{2}$ or $\pm\frac{1}{2}$ spin states to either a heavy or light hole is no longer valid at finite $k$.
            These issues were described in Winkler, 2003~\cite{Winkler2003}, and we will explore them further via spin-projected band structures in Sec.~\ref{section:bandstructure}.

        \subsubsection{Zeeman term}
            The Zeeman term for holes is described by~\cite{Winkler2003,Miserev2017,Rendell2021}
            \begin{align}
                \label{Eq:Hamiltonian_Zeeman}
                H_{\mathrm{Z}} = -2\kappa\mu^{}_{\mathrm{B}}\vec{B}\cdot\vec{J}
            \end{align}
            where $\kappa$ is an effective $g$ factor for holes, $\mu^{}_{\mathrm{B}}$ is the Bohr magneton, and $\vec{J}$ are the spin-$\frac{3}{2}$ spin matrices.
            The 2DHG we study in this paper is formed in a GaAs/Al\textsubscript{x}Ga\textsubscript{1-x}As heterostructure (Fig.~\ref{fig:2DHG_potentialwells}).
            Therefore we used $\kappa=1.2$, which is the literature value for bulk GaAs ~\cite{Winkler2003}.
            Since the Zeeman term is at least an order of magnitude smaller than the other terms in $H^{}_{\mathrm{2D}}$ for the magnetic strengths considered, we deem the $\kappa$ value for bulk GaAs to be a sufficient approximation.
            The final Hamiltonian is then the sum of Eq.~\ref{Eq:2DHG_Hamiltonian_2D} and Eq.~\ref{Eq:Hamiltonian_Zeeman}:
            \begin{align}
                \label{Eq:Hamiltonian_total}
                H = H^{}_{\mathrm{2D}} + H^{}_{\mathrm{Z}}
            \end{align}

        \subsubsection{Peierls substitution}
            Charged carriers in a magnetic field experience an in-plane Lorentz force and move in a cyclotronic motion.
            We approximate this behaviour by applying the Peierls substitution method, where a phase factor ($\phi$) called the Peierls phase is applied to the hopping terms ($t$) in our discretised Hamiltonian~\cite{Peierls1933, Peierls1997, Luttinger1951, Hofstader1976, Datta2012},
            $t \rightarrow te^{i\phi}$.
            This Peierls phase $\phi$ is given by:
            \begin{align}
                \label{Eq:Peierls_phase}
                \phi = \frac{ie}{\hbar}\int_{\vec{r}_{m}}^{\vec{r}_{n}}{\vec{A}\left( \vec{r} \right)} \cdot d\vec{r}
            \end{align}
            where $\vec{r}_{m}$ and $\vec{r}_{n}$ are the positions of the initial and final sites that the hopping term connects, and $\vec{A}$ is the magnetic vector potential:.   
            In this paper, we use the Landau gauge~\cite{Landau1930} in $x$ or $y$:
            \begin{align}
                \label{Eq:Landau}
                \vec{A} = -B^{}_{z}y\hat{x} \qquad \mathrm{or} \qquad \vec{A} = B^{}_{z}x\hat{y}
            \end{align}
            Gauge invariance was verified in our calculations by comparing the LDOS of devices rotated by 90 degrees to each other and ensuring that the results were identical.
            
    \subsection{Computational methods}
        \label{subsection:computational}
        We discretised Eq.~\ref{Eq:Hamiltonian_total} in real space using a central finite difference scheme ~\cite{Grossmann2007}, which allows the equation to be solved numerically.
        A discretisation grid size of 2~nm was chosen as the eigenvalues in our band structure calculations were sufficiently converged at this grid size.
        The open source Python package Kwant~\cite{Groth2014} was used for the numerical calculations throughout this paper.
        
        Conductance ($\mathcal{G}$) is calculated in the zero-bias and zero-temperature limit, which makes it directly proportional to the transmission function ($T$)~\cite{Datta2012, Lee2022}:
        \begin{align}
            \label{Eq:conductance}
            \mathcal{G}(E) &= \frac{e^{2}}{h} T(E)
        \end{align}
        We calculated the transmission function using the non-equilibrium Green's function (NEGF) formalism~\cite{Datta2012}.
        The details of the method and its implementation can be found in Lee et al., 2018~\cite{Lee2022} and Lee, 2023~\cite{Lee2023}.

        \begin{figure*}[htp]
            \centering
            \includegraphics[width=0.95\linewidth]{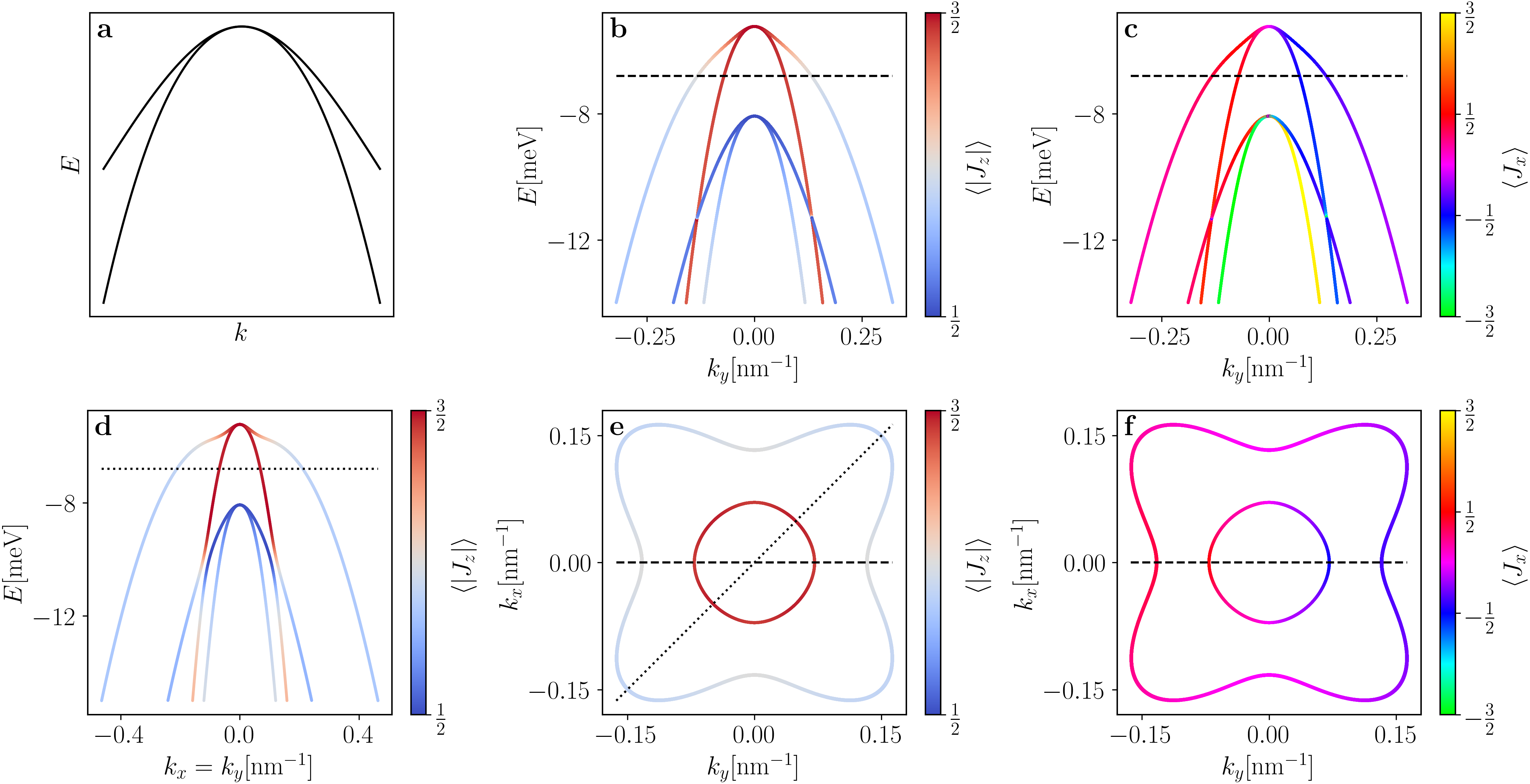}
            \caption{
            (a) Diagram of heavy hole subbands for an idealised 2DHG with Rashba spin-orbit coupling. 
            (b-f) Band structures calculated for a 2DHG in GaAs/Al\textsubscript{x}Ga\textsubscript{1-x}As.
            (b) and (c) are plotted along the $k^{}_{y}$ axis and the dashed lines here mark $1.6$~meV below the valence band edge, the energy at which (e) and (f) were calculated.
            (d) is plotted along the along the $k^{}_{x}=k^{}_{y}$ axis, and the dotted line marks $1.6$~meV below the valence band edge.
            (e) and (f) show the Fermi surface sliced at $1.6$~meV below the valence band edge and the dashed and dotted lines here indicate the $k^{}_{y}$ and $k^{}_{x}=k^{}_{y}$ axes respectively.
            The colour bars in (b), (d) and (e) indicate the expectation value of $\left| J^{}_{z} \right|$, whereas the colour bars in (c) and (f) indicate the expectation value of $J^{}_{x}$.
            }
            \label{fig:Bands_2D}
        \end{figure*}
   
        In our calculations, we take the wave functions corresponding to each available transmission mode in the injection lead at a chosen energy and calculate the probability density at each site in the scattering region to obtain the local density of states (LDOS)~\cite{Lee2023}:
        \begin{align}
            \label{Eq:LDOS_discrete}
            D^{}_{(l,w)} &= \sum^{}_{\alpha} \left\langle \psi^{}_{(l,w),\alpha} \middle| \psi^{}_{(l,w),\alpha} \right\rangle
        \end{align}
        where $(l,w)$ are the coordinates in the discretisation grid and $\alpha$ is an integer representing all possible transmission modes at the chosen energy.
        
        The band structures and Fermi surfaces shown in this paper were calculated based on Bloch's theorem~\cite{Ashcroft1976, kittel2004}.
        Spin projected versions of the conductance, LDOS, and band structures were obtained by applying a spin projection operator $\Hat{P}$ which is given by~\cite{Griffiths2018}:
        \begin{align}
            \label{Eq:projection_operator}
            \Hat{P} = | v \rangle \langle v |
        \end{align}
        where $|v\rangle$ is an eigenvector of the chosen spin operator.

        We chose the contact on the right side to be the injector as shown in Fig.~\ref{fig:TMF_system}, in order to compare to previous work on TMF performed in a 2DEG~\cite{Lee2022}.
        Therefore to focus holes from injector on the right to the collector on the left, the magnetic field is negative (points into the 2D plane) throughout this paper.  
        To confirm that the Onsager reciprocity relation~\cite{Jacquod2012} is respected, we also performed test calculations where the injector and collector are swapped, and the magnetic field direction is reversed. 
        As expected the results were identical within the numerical noise limits of the calculations.
    
\section{Band structure of 2D holes in an unbounded quantum well}
\label{section:bandstructure}
    To understand the dynamics of charge transport in 2DHGs, we first consider the band structure in the vicinity of the valence band maximum.
    In a 2DHG, the four-fold degeneracy in the valence band is lifted due to quantum confinement effects, and splits into two two-fold degenerate bands usually referred to as the heavy hole (HH) and light hole (LH) bands~\cite{Winkler2002}.
    In 2DHG experiments where the charge density is small, only the highest valence subbands, i.e.~the HH states, would be occupied.
    It is then common to describe the dispersion relation of the HH subbands for a 2DHG with Rashba spin-orbit coupling using an equation of the form $E = ak^2 \pm bk^3$~\cite{Winkler2000, Bladwell2015, Marcellina2017, Liu2018, Rendell2023}.
    The HH bands would look like Fig.~\ref{fig:Bands_2D}a, where each subband corresponds to a pseudospin state labelled HH$+$ or HH$-$.        
    We calculated the HH-LH subbands for a 2DHG within a GaAs/Al\textsubscript{x}Ga\textsubscript{1-x}As heterostructure using the 4$\times$4 Luttinger Hamiltonian described in section~\ref{section:Methods}.
    In Figs.~\ref{fig:Bands_2D}b-f, we plot the band structures along different $k$-axes and with various spin projection schemes applied.

    Fig.~\ref{fig:Bands_2D}b shows the band structure plotted along the $k^{}_{y}$ axis ($k^{}_{x}=0$) while in Fig.~\ref{fig:Bands_2D}d the band structure is plotted along the $k^{}_{x}=k^{}_{y}$ axis.
    The colours in both figures represent the expectation value of $|J^{}_{z}|$  at each point, where red corresponds to $ J_z =\pm\frac{3}{2}$ states (HH) and blue corresponds to $ J_z =\pm\frac{1}{2}$ states (LH).
    We see that as expected, the HH and LH subbands are well separated at the $\Gamma$-point $(k=0)$ in Fig.~\ref{fig:Bands_2D}b.
    However, significant mixing occurs between the bands as we move away from the $\Gamma$-point.
    When the band structure is projected into individual $J^{}_{z}=+\frac{3}{2}, -\frac{3}{2}, +\frac{1}{2}, -\frac{1}{2}$ states, there is a significant component of each state present in all the subbands.
    Therefore, at finite $k$, the states in each subband are a superposition of all HH and LH states, and cannot be clearly labelled as corresponding to any single pure $J_z$ state.

    Comparing Fig.~\ref{fig:Bands_2D}b and Fig.~\ref{fig:Bands_2D}d, we also note that certain features in the band structure change significantly depending on the axis it is plotted along, most notably the curvature of the highest subband, and the presence of an anti-crossing between the second and third subbands.
    When we plot the band structure at a fixed energy, chosen to be $1.6$~meV below the valence band edge in Fig.~\ref{fig:Bands_2D}e, we observe that the first (outer) and second (inner) highest subbands do not appear as concentric circles as the dispersion relation used in Fig.~\ref{fig:Bands_2D}a would suggest they should.
    Instead, the outer subband is shaped like a wavy squircle (i.e.~a superellipse modified by a sinusoidal function) which resembles a piece of toast.
    Since this band structure displays the allowed momentum states at a certain energy, the shape of the subbands in $k$-space is related to the shape of the hole's path in real space when an external magnetic field is applied~\cite{Ashcroft1976}.
    This suggests that a hole injected into a TMF device with a $k$ value in the outer band would travel in a toast-shaped trajectory.
    This can also be interpreted as the holes in a 2DHG having an anisotropic in-plane effective mass.

    Figs.~\ref{fig:Bands_2D}c and~\ref{fig:Bands_2D}f show the same band structures as in Figs.~\ref{fig:Bands_2D}b and~\ref{fig:Bands_2D}e respectively, but with spin projections for the expectation values of $J^{}_{x}$.
    The two subbands that result from the lifting of the degeneracy of the HH band are referred to as ``spin-split'' subbands in the literature~\cite{Winkler2003}.
    The effective two-state Hamiltonians commonly used to describe Rashba SOI in 2DHGs usually couples momentum with spin in perpendicular in-plane directions (i.e.~$k_y$ with $\sigma_x$ and $k_x$ with $\sigma_y$)~\cite{Winkler2003, Bladwell2015, Marcellina2017, Liu2018, Rendell2023}, similar to 2DEGs.
    Therefore, we would expect spin behaviour in the HH subbands to be similar to 2DEG band structures~\cite{Lee2022}, where the inner and outer subbands have opposing spin chiralities.
    This means that at any particular energy, when the band structure is spin projected along the $x$ or $y$ axis, the inner and outer subbands should appear to be anti-phased, i.e.~ when the inner subband is spin up, the outer subband would be spin down, and vice versa.
    However, we can see in Figs.~\ref{fig:Bands_2D}c and~\ref{fig:Bands_2D}f that the highest two subbands are not anti-phased in spin.
    Instead, they appear to be spin-momentum locked at finite $k$, i.e.~$\langle J^{}_{x} \rangle$ is positive or negative when $k^{}_{y}$ is negative or positive respectively.  

    \begin{figure}[htp]
        \centering
        \includegraphics[width=0.9\linewidth]{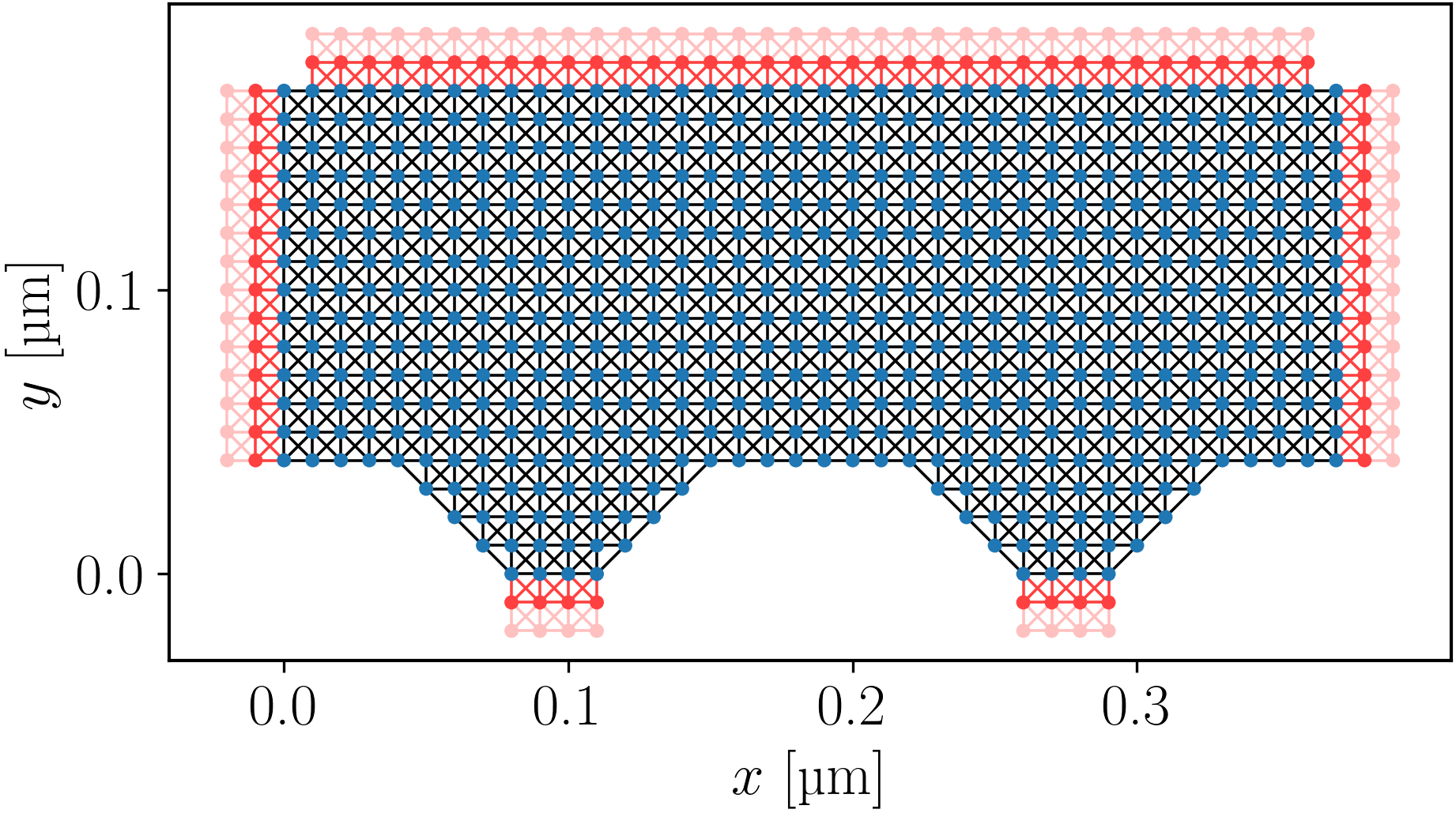}   
        \caption{
        Example diagram of a discretised grid used to model a TMF device with 50~nm wide injector and collector leads separated by a distance of 150~nm.
        The Hamiltonian is discretised in real space, giving onsite terms shown as blue dots, and hopping terms shown as solid black lines that connect neighbouring sites.
        Semi-infinite leads are indicated by the red dots.
        Here the grid spacing is set to be 10~nm for simplicity.
        In our calculations, the grid spacing is set to 2~nm and the separation between the leads is much larger as seen in Fig.~\ref{fig:LDOS}.
        }
        \label{fig:TMF_grid}
    \end{figure}
    
    While not shown here, the same band structures projected in $J^{}_{y}$ are identical to the band structures for $J^{}_{x}$ but with the colours rotated by $90^{\circ}$ around the $k^{}_{z}$-axis.   
    As was the case with the $J^{}_{z}$-projected band structures, when we projected the band structure into individual $ J^{}_{x} = +\frac{3}{2}, -\frac{3}{2}, +\frac{1}{2}, -\frac{1}{2}$ states, all the subbands correspond to superpositions of each of the $J^{}_{x}$ states.
    When we look at just the $J^{}_{x} = \pm\frac{3}{2}$ or the $J^{}_{x} = \pm\frac{1}{2}$ states, the band structure appears anti-phased in spin (See Appendix~\ref{Appendix:Jx_bandstructures}).
    It is only when all four states are considered that the band structure stops being anti-phased in spin.
    This shows that a typical two-state effective model is insufficient to describe the spin behaviour in a GaAs/Al\textsubscript{x}Ga\textsubscript{1-x}As 2DHG, even when assuming only the HH subbands are occupied.
    We emphasise here that the discussions in this paper pertain to 2DHGs in GaAs/Al\textsubscript{x}Ga\textsubscript{1-x}As single heterojunctions where the perpendicular confinement at the interface is as shown in Fig.~\ref{fig:2DHG_potentialwells}.
    There are other types of heterostructures where a two-state effective model can be applied (see Appendix~\ref{Appendix:narrow_wells}) and one would expect spin-polarised transport which is more analogous to that seen in 2DEG models with Rashba.
    Understanding that the eigenstates in the band structure of a 2DHG are superpositions of different spin projections is key in interpreting results from transverse magnetic focusing experiments.

\section{Transverse magnetic focusing in an idealised device}
\label{section:hardwall}  
    \begin{figure}[htp]
        \centering
        \includegraphics[width=0.9\linewidth]{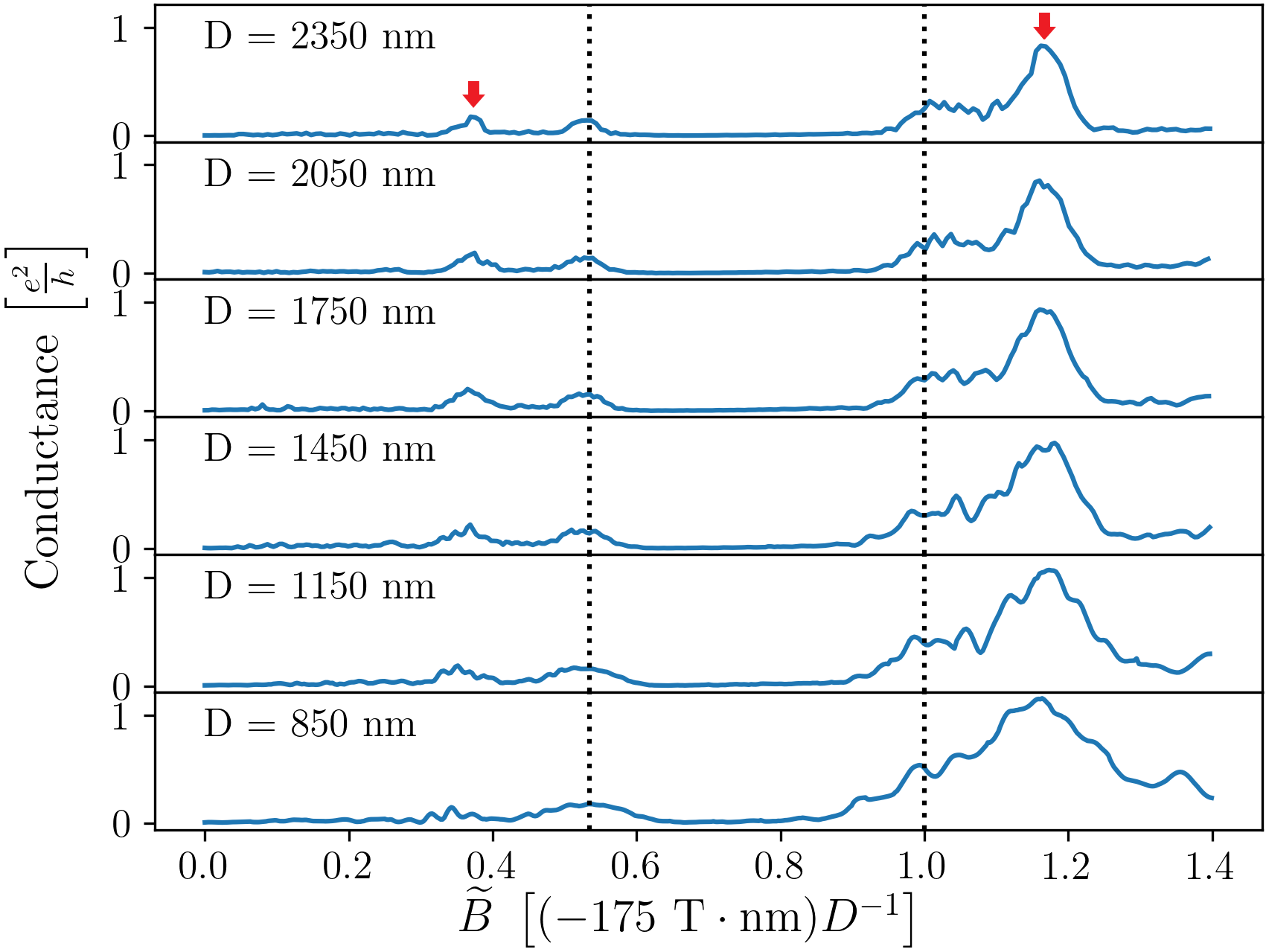}
        \caption{
        Conductance spectra calculated for TMF devices with 50~nm wide injector and collector leads separated by distances ranging from 850~nm to 2350~nm.
        The Fermi energy was set to $1.6$~meV below the valence band edge in each calculation.
        The dotted lines indicate the first and second focusing magnetic field strengths corresponding to $k^{}_{y}$ = 0.07~nm$^{-1}$ and $k^{}_{y}$ = 0.13~nm$^{-1}$ respectively, calculated from Eq.~\ref{Eq:larmor}.
        The perpendicular magnetic field strengths are normalised to the second focusing field strengths for each $D$, such that {$B_{z} = \widetilde{B} \times D^{-1} \times (-175~\mathrm{T}\cdot\mathrm{nm})$}.        
        }
        \label{fig:TvB_varyD}
    \end{figure}    
    
    By discretising the Hamiltonian (Eq.~\ref{Eq:Hamiltonian_total}) in real space, we numerically modelled a TMF device with open boundaries on three sides, and a hard wall boundary on the bottom as illustrated by the discretisation grid in Fig.~\ref{fig:TMF_grid}.
    Two 50~nm wide semi-infinite leads were attached to the bottom hard wall boundary to serve as the injector and collector in our calculations.    
    We chose to set the Fermi energy in our calculations to be $1.6$~meV from the valence band edge, approximately in the middle of the gap between the HH and LH subbands as can be seen in Fig.~\ref{fig:Bands_2D}b.
    This energy is similar to the Fermi energy in the single heterojunction experiments performed by Rendell et al.~\cite{Rendell2015, Rendell2022, Rendell2023}, where the carrier density is $1.89 \times 10^{11}$~cm$^{-2}$.
    Looking at the band structure in Fig.~\ref{fig:Bands_2D}b, two positive $k^{}_{y}$ values (approximately 0.07~nm$^{-1}$ and 0.13~nm$^{-1}$) correspond to an energy of $1.6$~meV below the valence band edge.
    Substituting these $k^{}_{y}$ values into Eq.~\ref{Eq:larmor} gives the focusing magnetic field strengths where peaks would be expected in the conductance spectra.

    Fig.~\ref{fig:TvB_varyD} shows the conductance spectra calculated for TMF devices with the distance between the injector and collector leads ($D$) varied from 850~nm to 2350~nm.
    The first and second focusing fields corresponding to {$k^{}_{y}$ = 0.07~nm$^{-1}$} and {$k^{}_{y}$ = 0.13~nm$^{-1}$} respectively were calculated from Eq.~\ref{Eq:larmor} for each $D$ value and are shown as dashed lines.
    As $D$ is changed, the magnetic field strengths at which peaks occur in the conductance spectrum shifts as well.
    The magnetic field strengths are expressed as a ratio of the second focusing field strengths ($\widetilde{B}$), {i.e.~$B_{z} = \widetilde{B} \times D^{-1} \times (-175~\mathrm{T}\cdot\mathrm{nm})$}.

    \begin{figure}[htp]
        \centering
        \includegraphics[width=0.9\linewidth]{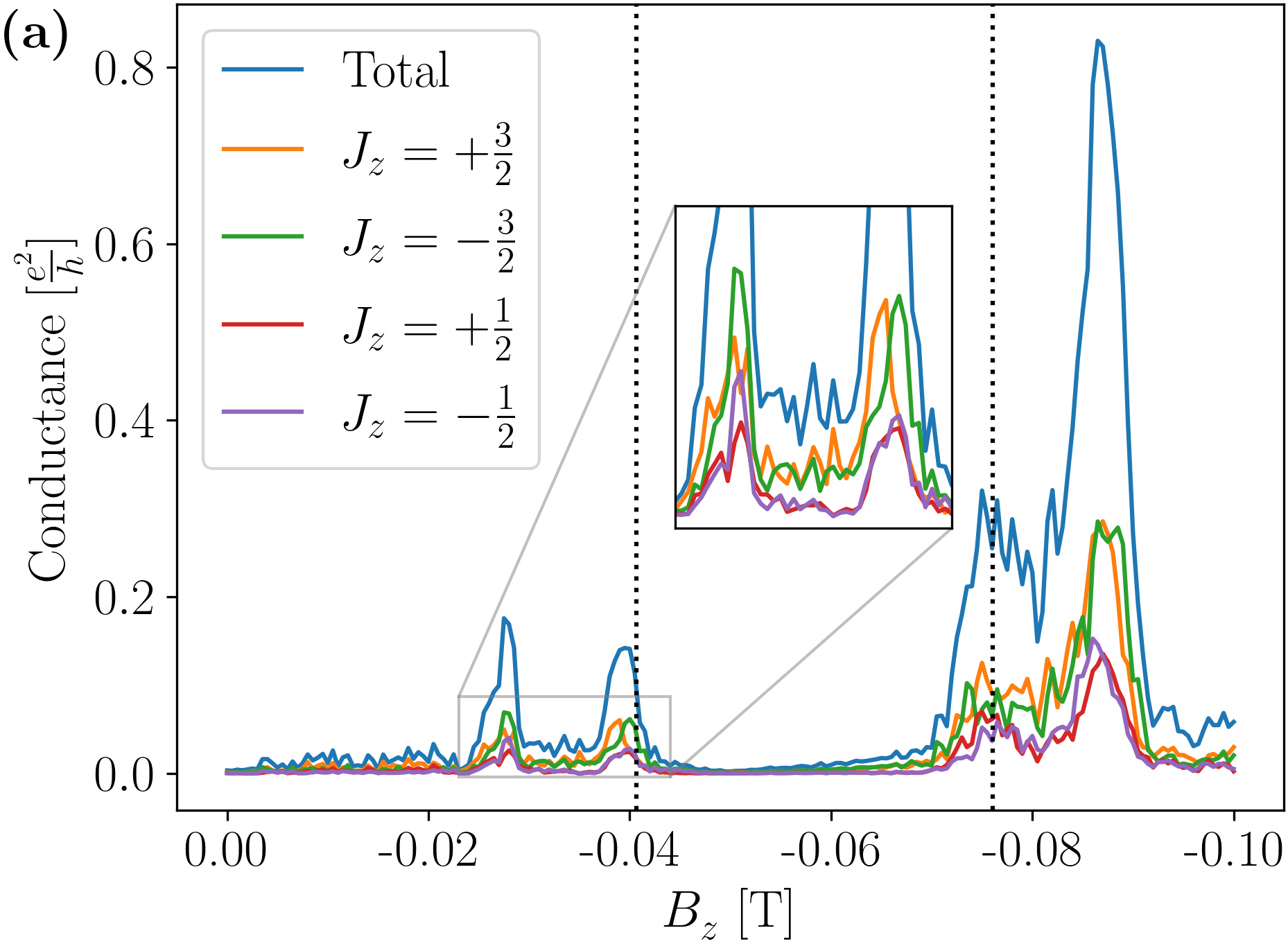}
        \includegraphics[width=0.9\linewidth]{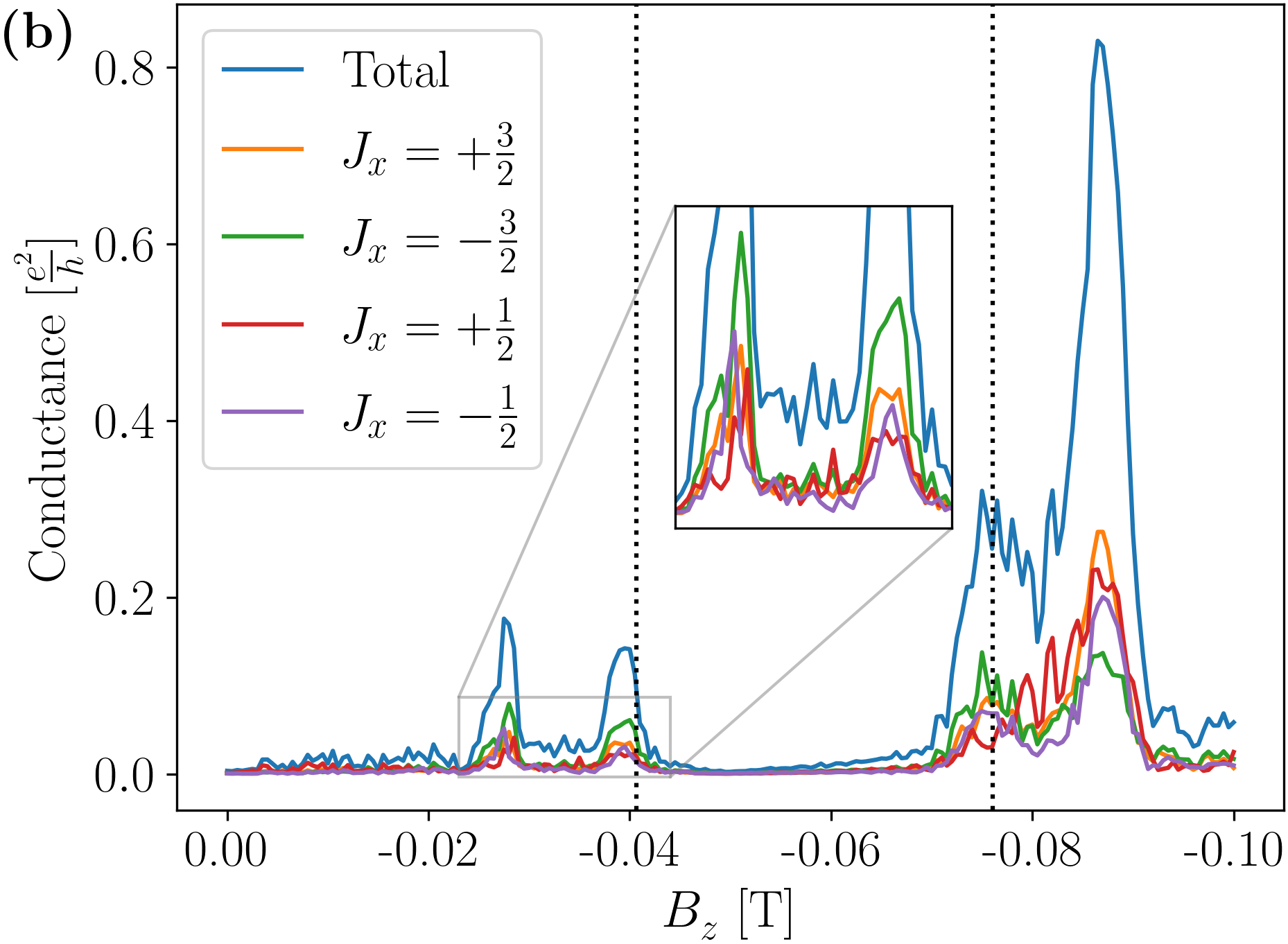}
        \caption{
        Conductance spectra for a TMF device with 50~nm wide leads separated by 2350~nm, spin projected in (a) $J^{}_{z}$ and (b) $J^{}_{x}$.
        The dotted lines indicate the first and second focusing magnetic field strengths corresponding to $k^{}_{y}$ = 0.07~nm$^{-1}$ and $k^{}_{y}$ = 0.13~nm$^{-1}$ respectively, calculated from Eq.~\ref{Eq:larmor}.
        The insets show an enlarged view of the boxed area around the first two peaks.
        }
        \label{fig:TvB_2300nm}
    \end{figure}
    
    We see that for all $D$ values, a peak is observed in the conductance spectra at or near the expected focusing field strengths.
    However, we note that the first expected focusing peaks all have a paired peak at a weaker $B^{}_{z}$ while the second expected focusing peaks have a paired peak positioned at a stronger $B^{}_{z}$, as indicated by the red arrows.
    As $D$ becomes shorter, the peaks become wider and begin to overlap, and eventually a peak pair can present as a single wide peak, e.g.~the second focusing peak in the spectrum for $D=850$~nm.
    The shape of the peaks and the gap between the peaks within each peak pair remain consistent as $D$ is changed, suggesting that these extra peaks are not the product of interference, but their origin remains unclear.
    While a similar double paired peaks structure has been experimentally measured before~\cite{Rendell2015}, we do not see the same structure when we explore this further with a more realistic simulation in Sec.~\ref{section:QPC}, therefore these peaks may be due to a mode mismatch at the interface between the semi-infinite leads and the scattering area.

    \begin{figure}[htp]
        \centering
        \includegraphics[width=0.9\linewidth]{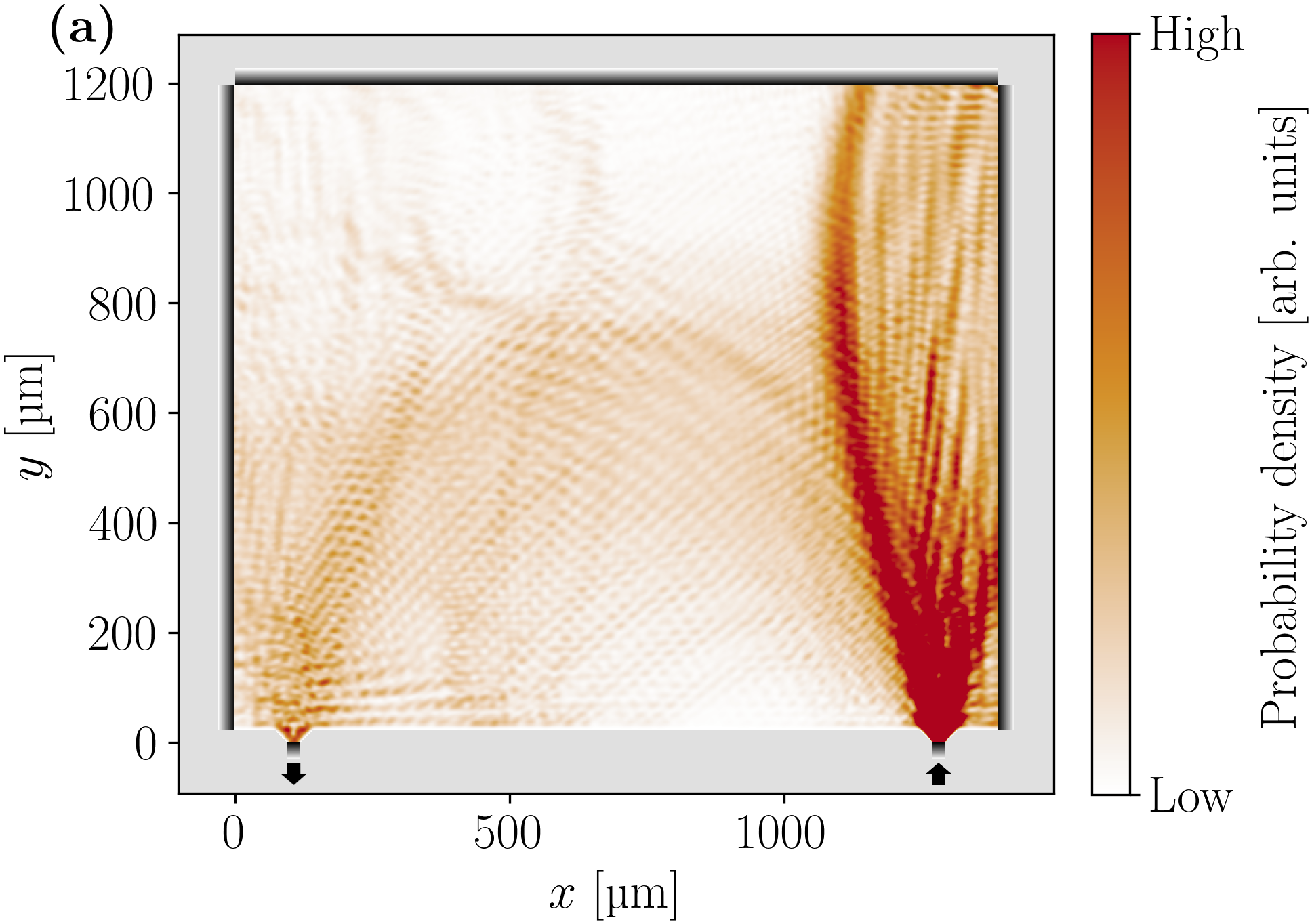}
        \includegraphics[width=0.9\linewidth]{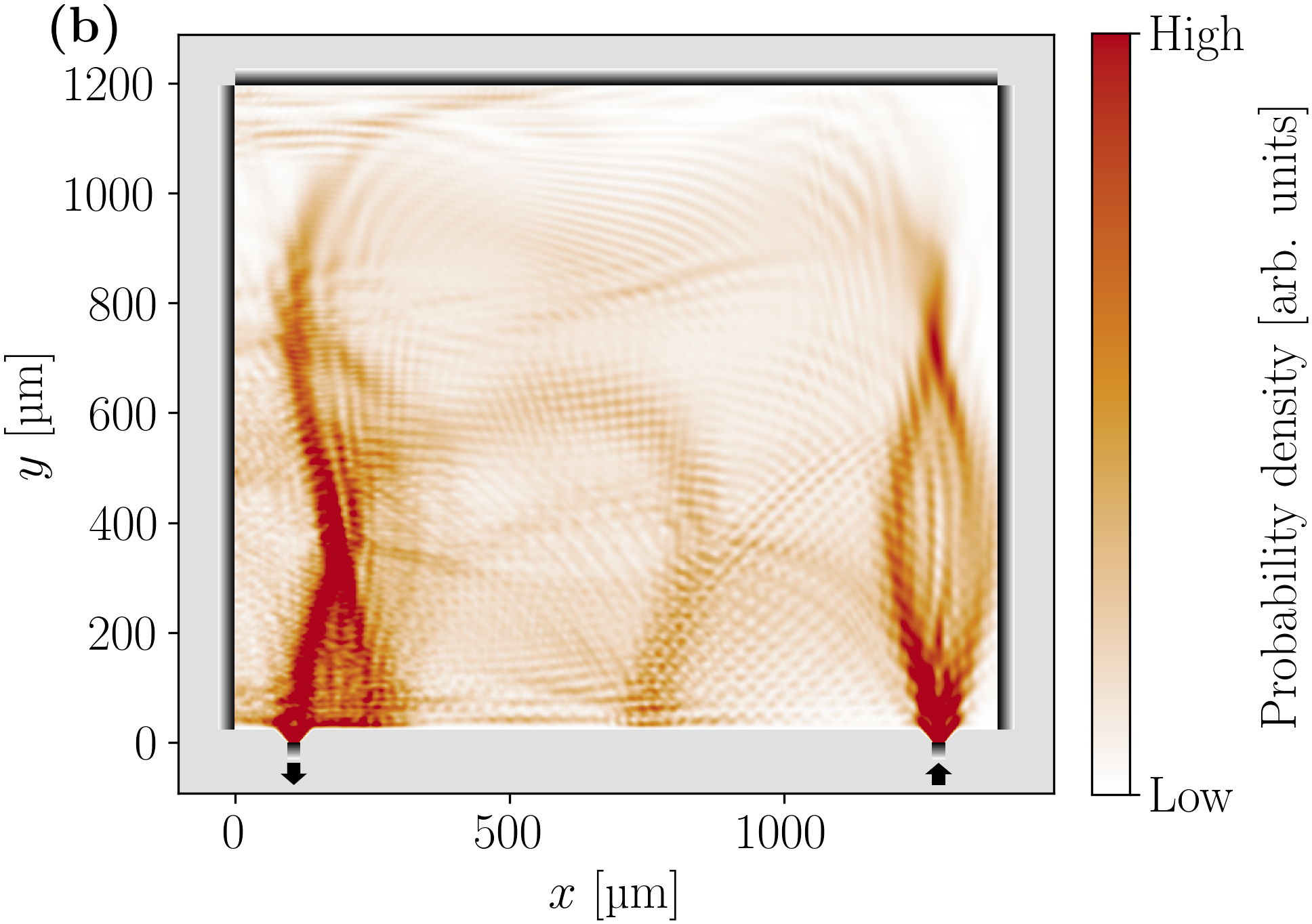}
        \caption{
        LDOS for a TMF device with 50~nm wide leads separated by 2350~nm, calculated at magnetic field strengths of (a) $-0.0394$~T and (b) $-0.0870$~T.
        The colour bar shows an arbitrarily scaled probability density, where the darker the colour, the higher the density at that point.
        }
        \label{fig:LDOS}
    \end{figure}
    
    We now focus on the conductance spectrum for $D=2350$~nm and calculate spin projections in $J^{}_{z}$ and $J^{}_{x}$ to better understand the spin compositions at each of the peaks.
    The orange, red, green and purple lines in Fig.~\ref{fig:TvB_2300nm}a show the contributions to the total conductance from each of the $J^{}_{z} = +\frac{3}{2}, -\frac{3}{2}, +\frac{1}{2}$, and $-\frac{1}{2}$ states respectively.
    Similarly, the coloured lines in Fig.~\ref{fig:TvB_2300nm}b show the contributions from the $J^{}_{x}$ states.
    Given a spectrum showing only the total conductance, a common interpretation based on the band structure in Fig.~\ref{fig:Bands_2D} and Eq.~\ref{Eq:larmor} would be that the peaks found at the first and second expected focusing field strengths are spin-polarised, i.e.~each peak would correspond to a different spin when projected along the same axis.
    However, we can see in Fig.~\ref{fig:TvB_2300nm} that the peaks are not spin-polarised in either $J^{}_{z}$ or $J^{}_{x}$.

    \begin{figure}[htp]
        \centering
        \includegraphics[width=0.9\linewidth]{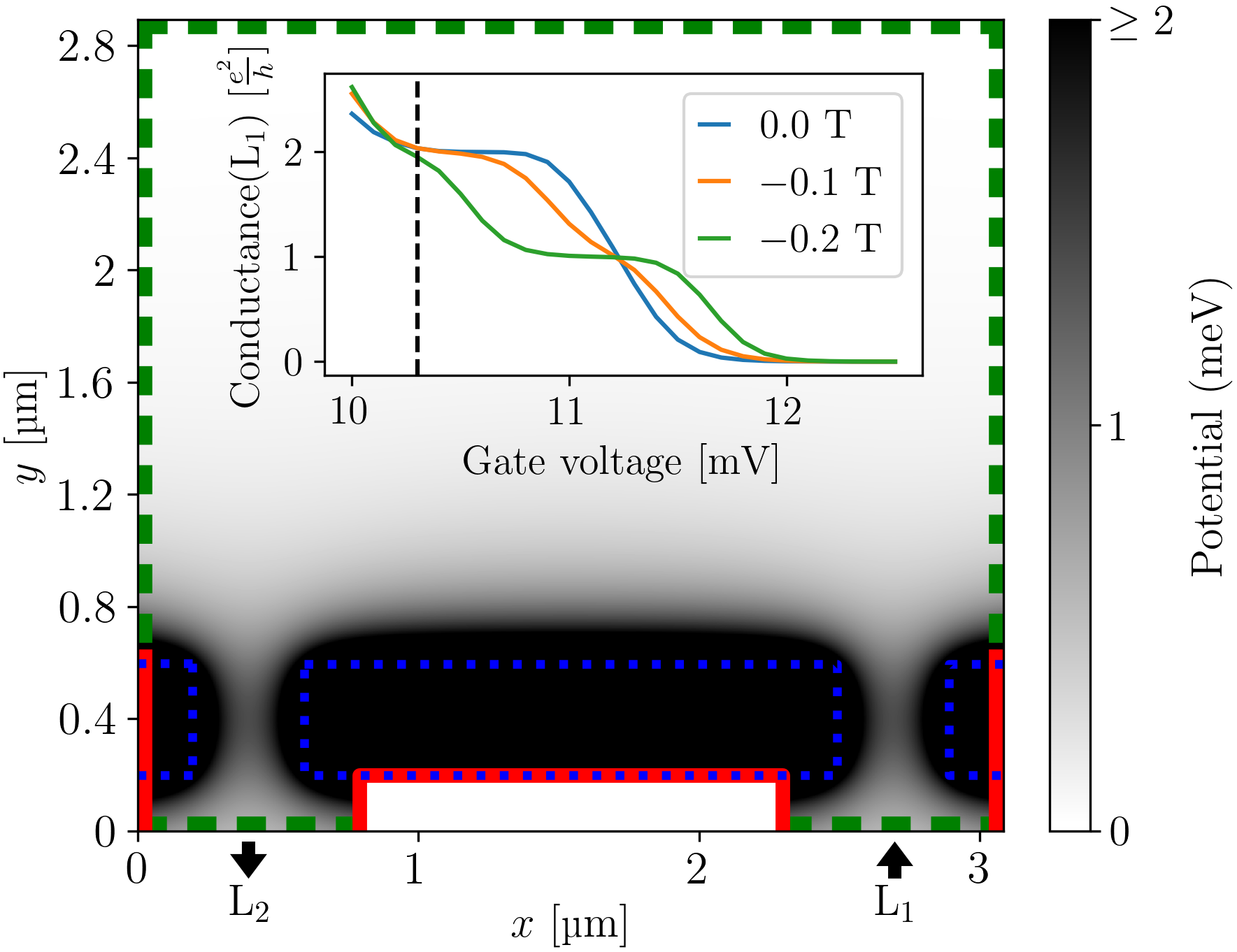}
        \caption{
        Diagram of setup used to model a TMF device with QPC leads.
        Red solid lines indicate closed boundaries.
        Green dashed lines indicate open boundaries where semi-infinite leads are attached, through which holes can enter or exit the system.
        L$_{1}$ and L$_{2}$ indicate the leads used for injection and collection respectively in our calculations.
        Blue dotted lines indicate where metal gates are placed above the 2DHG.
        The colour bar shows the potential in the 2DHG plane due to a voltage of 10.3~mV applied at the gates.
        This colour bar is cut off at 2~meV to better display the QPCs formed by the gates.
        (inset) Total conductance from $L^{}_{1}$ to all other leads as a function of the voltage applied to the gates, calculated for $B^{}_{z}$ = $0.0$~T, $-0.1$~T and $-0.2$~T.
        The dashed line marks the gate voltage used (10.3~mV) for the potential shown in the main figure.
        }
        \label{fig:QPC_potential}
    \end{figure}
    
    Finally, we visualise the distribution of the holes in the TMF device by calculating the LDOS.
    Fig.~\ref{fig:LDOS}a and Fig.~\ref{fig:LDOS}b show the LDOS when $B^{}_{z}=-0.0394$~T and $-0.0870$~T respectively.
    These $B^{}_{z}$ values correspond to the second and fourth peaks in Fig.~\ref{fig:TvB_2300nm}.
    Fig.~\ref{fig:LDOS}a shows that holes with smaller $k$ (since $B^{}_{z}$ is smaller) travel a circular path whereas holes with larger $k$ travel in a toast-shaped trajectory as shown in Fig.~\ref{fig:LDOS}b, qualitatively agreeing with the shapes of the inner and outer subbands in Fig.~\ref{fig:Bands_2D}f.
    The LDOS in Fig.~\ref{fig:LDOS}b also shows that there is a higher probability of finding the hole in the outer trajectory compared to the inner trajectory.
    This is reflected in the conductance spectra (Figs~\ref{fig:TvB_varyD} and~\ref{fig:TvB_2300nm}), where we saw that the peaks at the first focusing field strength are of a lower intensity than the peaks at the second focusing field strength.
    There are also additional structures seen in the LDOS plots that do not have corresponding features in the band structure, such as the branching trajectories near the injection and collection points.
    
    We note here that the spherical approximation is sometimes used when obtaining analytical solutions to the Luttinger Hamiltonian (Eq.~\ref{Eq:2DHG_LuttingerH}), where the Luttinger parameters are changed such that $\gamma^{}_{2}=\gamma^{}_{3}$.
    Using our approach, we can also calculate the conductance spectrum and LDOS using this spherical approximation, in which case there are only two peaks in the conductance spectrum, there is no branching structure in the LDOS, and all the trajectories in the LDOS are circular.
    Therefore it is important to carefully consider the approximations taken when interpreting results from 2DHG TMF experiments so that important features are not hidden. 
    In fact, in the next section we will show that having an accurate description of the QPCs is essential for obtaining an accurate response. 
    In section~\ref{section:QPC} we also observe only two peaks in the conductance spectrum, however this is due to \emph{increasing} the accuracy of the description (rather than decreasing the accuracy via the spherical approximation).

\section{The influence of quantum point contacts}
\label{section:QPC}
    In the previous section, we modelled a TMF device with semi-infinite leads connected directly to the scattering area.
    This approach allows easy and independent control of physical parameters such as the width and positioning of the leads~\cite{Lee2022}.
    However, these semi-infinite leads essentially describe a narrow wire confined by an infinite square well perpendicular to the transport direction, whereas the scattering area is better described as an infinite 2DHG plane.
    Due to the effects of quantum confinement, the band structure of a semi-infinite lead differs significantly from the band structure of the scattering area (Fig.~\ref{fig:Bands_2D}).
    Therefore, there is a mismatch in band structures at the interface between a semi-infinite lead and the scattering area.
    This mismatch could be the cause of the extra peaks present in the conductance spectra shown in Figs.~\ref{fig:TvB_varyD}~and~\ref{fig:TvB_2300nm}.

    In this section we explore the interface mismatch between the leads and the scattering area by modifying our model to more closely match experimental setups.
    In experiments, the leads are quantum point contacts (QPCs) formed by applying voltage to surface gates placed above the 2DHG.
    We model this as shown in Fig.~\ref{fig:QPC_potential}, where the green dashed lines indicate open boundaries and the solid red lines indicate closed boundaries.
    The blue dotted lines indicate where surface gates are placed 122~nm above the 2DHG plane.
    The gates on the bottom left and bottom right of the diagram extend 3~$\unit{\um}$ beyond the left and right boundaries respectively.

    The electric potential generated within the 2DHG plane by voltage applied to the gates is approximated using the method derived by Davies et al.~\cite{Davies1995} and then added to the Hamiltonian (Eq.~\ref{Eq:Hamiltonian_total}).
    This potential forms QPCs between the scattering area and the semi-infinite leads used for injection ($L^{}_{1}$) and collection ($L^{}_{2}$) as seen in Fig.~\ref{fig:QPC_potential}, allowing for a smoother transformation between them.
    The leads here are also much wider (800~nm) compared to the leads used in Sec.~\ref{section:hardwall} (50~nm), thus the effects of quantum confinement in the leads is reduced, which lessens the mismatch at the interface.
    
    With the Fermi energy set to $1.6$~meV below the valence band edge, we calculated the total transmission from the lead $L^{}_{1}$ to all other leads in the system while varying the gate voltage and perpendicular magnetic field.
    This gives us the conductance of the right QPC in Fig.~\ref{fig:QPC_potential}, which is plotted against the gate voltage for $B^{}_{z}$ = 0.0~T, $-0.1$~T and $-0.2$~T in the inset figure of Fig.~\ref{fig:QPC_potential}.
    In our calculations, we chose to use a gate voltage of 10.3~mV since the conductance is approximately one conductance quantum ($2e^2/h$) at this voltage for magnetic field strengths of interest to us ($B^{}_{z} > -0.2$~T).

    \begin{figure}[htp]
        \centering
        \includegraphics[width=0.9\linewidth]{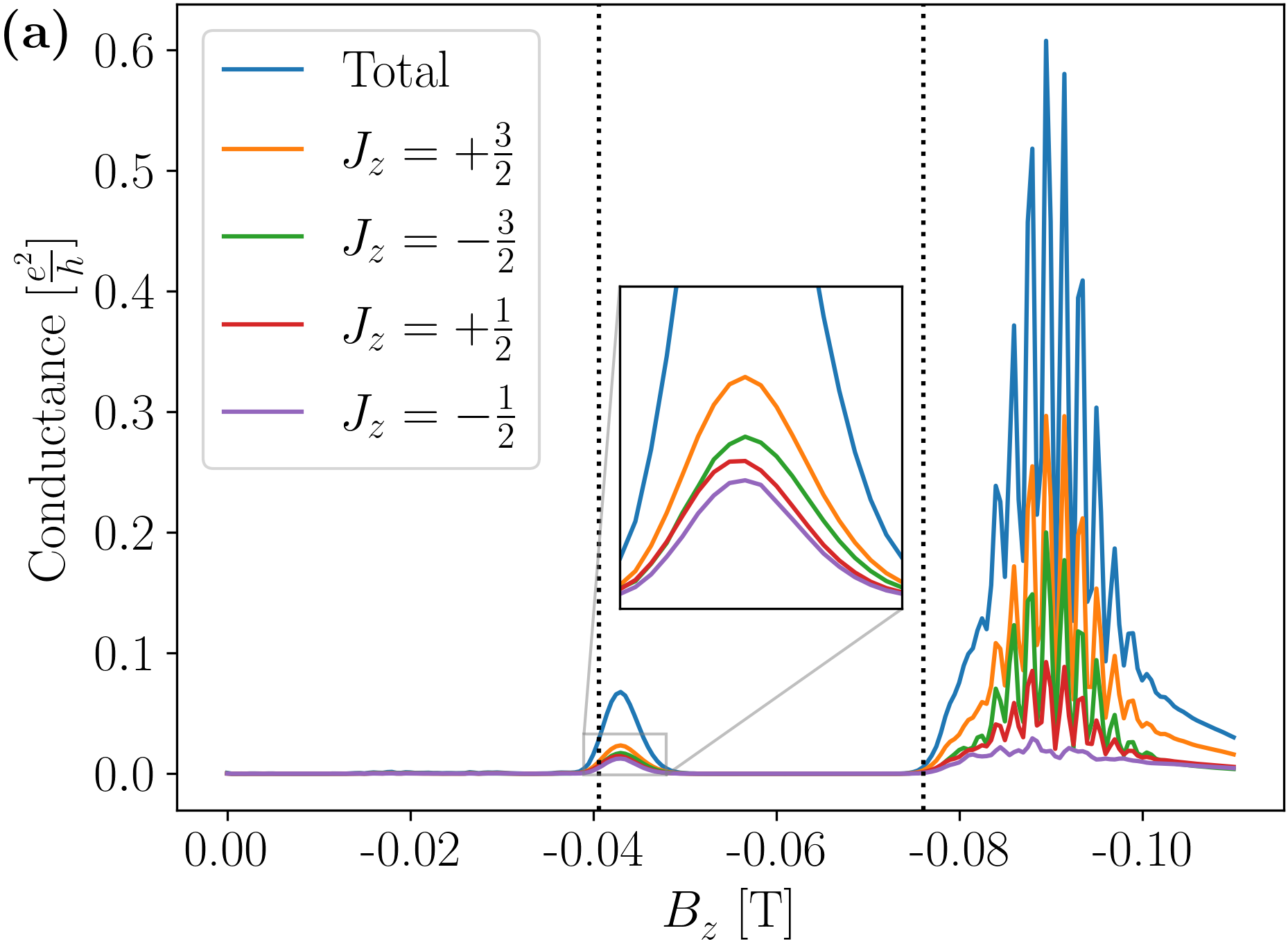}
        \includegraphics[width=0.9\linewidth]{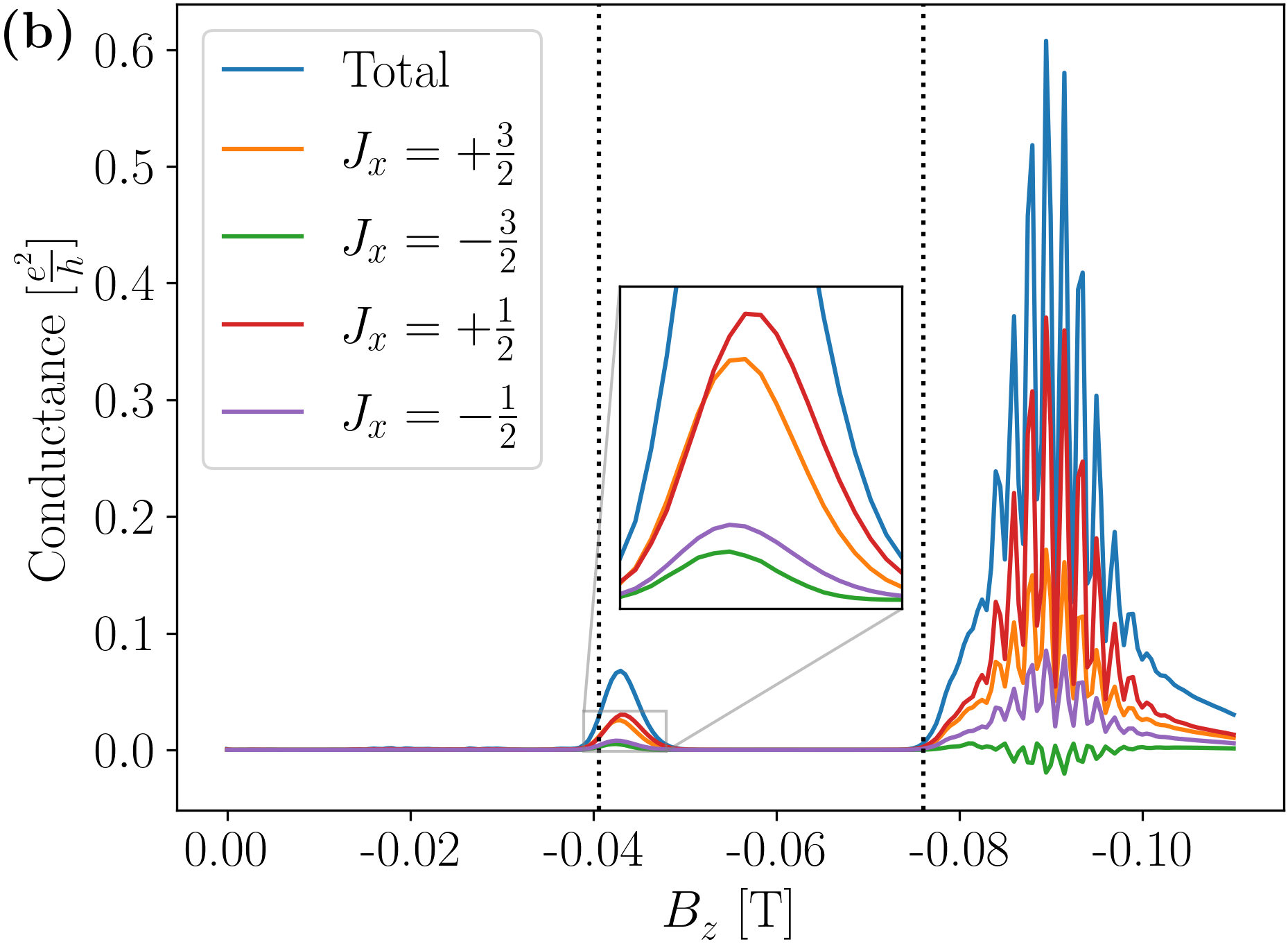}
        \caption{
        Conductance spectra for a TMF device modelled with QPC leads (Fig.~\ref{fig:QPC_potential}) separated by 2300~nm, spin projected in (a) $J^{}_{z}$ and (b) $J^{}_{x}$.
        The top gate voltage was set to 10.3~mV, such that the conductance through the injection QPC was one conductance quantum.
        The Fermi energy was set to be $1.6$~meV below the valence band edge.
        The dotted lines indicate the expected focusing magnetic field strengths for $k$-values corresponding to this Fermi energy.
        The insets show an enlarged view of the boxed area around the first peak.
        }
        \label{fig:QPC_TvB}
    \end{figure}

    Fig.~\ref{fig:QPC_TvB} shows the conductance spectra calculated for a TMF device modelled with QPCs formed by applying a voltage of 10.3~meV to surface gates.
    The distance between the two QPCs, measured from the centres of each of the gaps between the gates, was set to 2300~nm.
    Since the band structure of the scattering area remains the same as it was in Sec.~\ref{section:hardwall}, we again use $k^{}_{y}=0.07$~nm$^{-1}$ and 0.13~nm$^{-1}$ to calculate the expected focusing magnetic fields from Eq.~\ref{Eq:larmor}.
    The focusing fields are $-0.0407$~T and $-0.0761$~T and are indicated by the dotted lines in Fig.~\ref{fig:QPC_TvB}.
    We see in Fig.~\ref{fig:QPC_TvB}(a) that there are two peaks in the spectrum, centred at approximately $B^{}_{z}=-0.0430$~T and $B^{}_{z}=-0.0895$~T, both at slightly stronger magnetic fields than the corresponding expected focusing fields respectively.
    The peaks observed at $B^{}_{z}=-0.0276$~T and $B^{}_{z}=-0.0750$~T in Fig.~\ref{fig:TvB_2300nm} are not present here.

\begin{figure}[htp]
        \centering
        \includegraphics[width=0.9\linewidth]{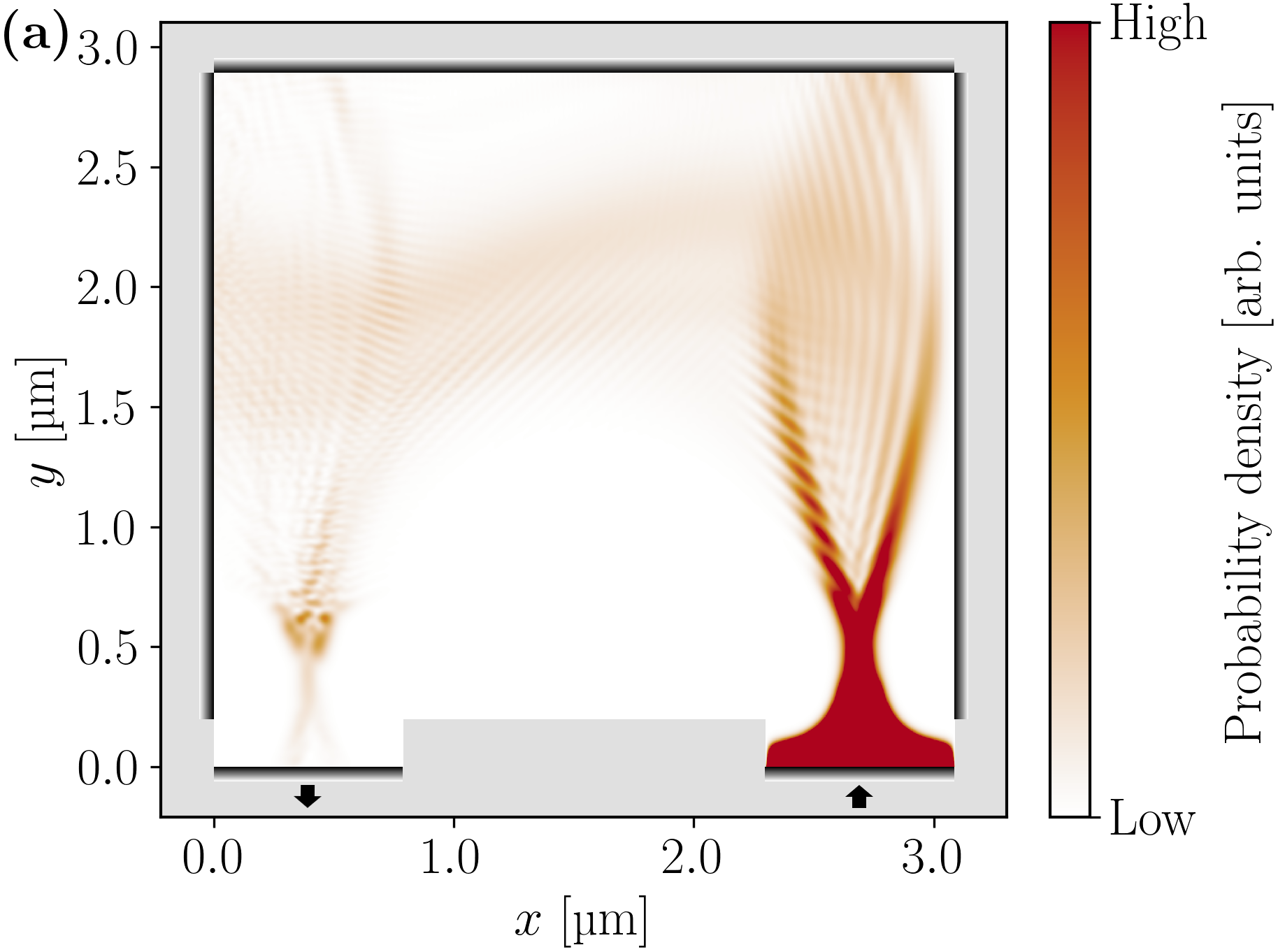}
        \includegraphics[width=0.9\linewidth]{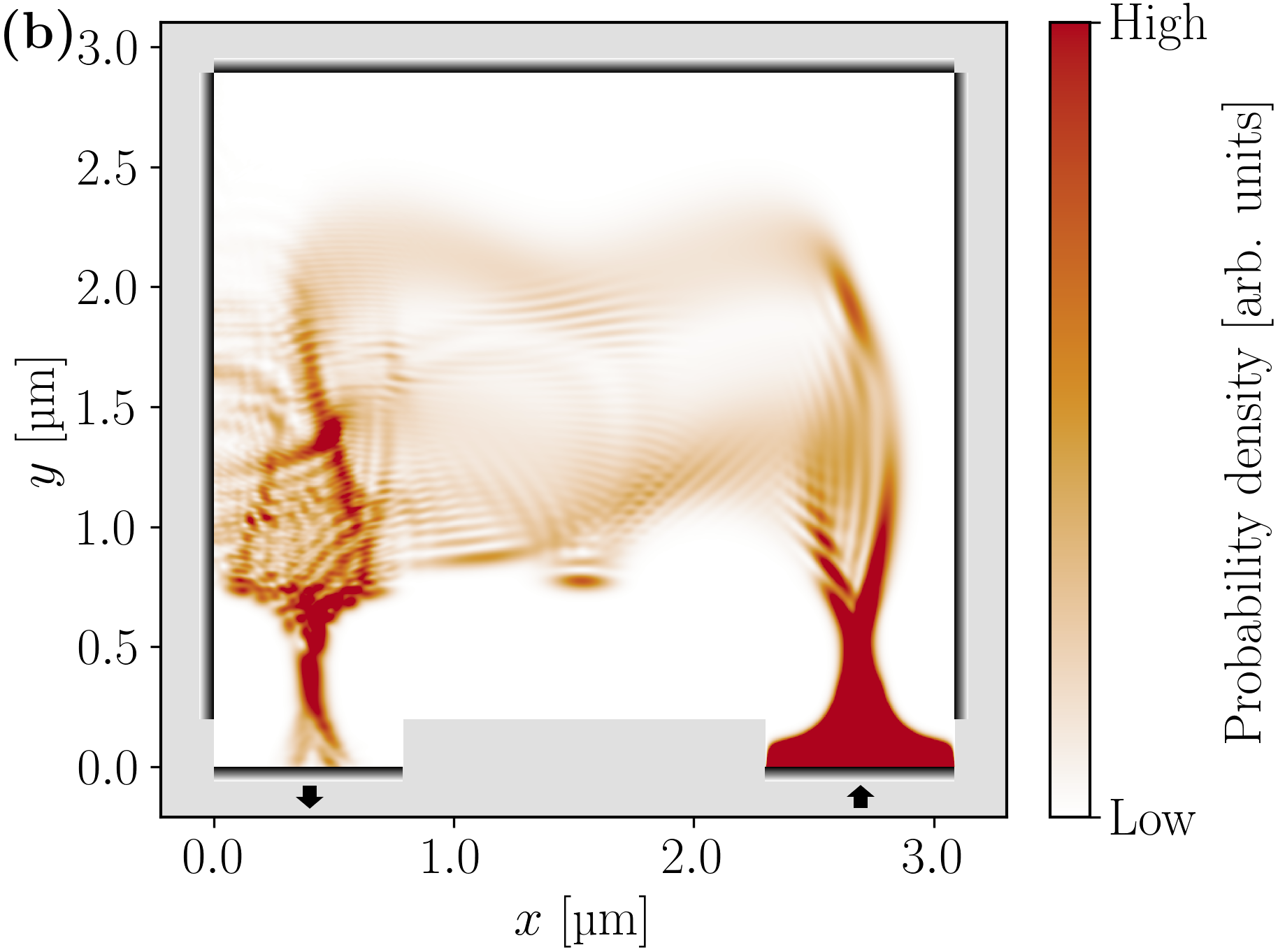}
        \caption{
        LDOS for a TMF device modelled with QPC leads (Fig.~\ref{fig:QPC_potential}) separated by 2300~nm, calculated for (a)~$B^{}_{z} = -0.0430$~T and (b)~$B^{}_{z} = -0.0895$~T.
        The top gate voltage was set to 10.3~mV, such that the conductance through the injection QPC was one conductance quantum.
        The Fermi energy was set to be $1.6$~meV below the valence band edge.
        }
        \label{fig:QPC_LDOS}
    \end{figure}

    When we calculate the LDOS for magnetic field strengths corresponding to the peak positions as shown in Fig.~\ref{fig:QPC_LDOS}, we notice a general similarity to the LDOS plots in Fig.~\ref{fig:LDOS}.
    The LDOS corresponding to the first peak (Fig.~\ref{fig:QPC_LDOS}a) shows a faint circular trajectory and the LDOS corresponding to the second peak (Fig.~\ref{fig:QPC_LDOS}b) shows a toast-shaped trajectory, matching the shapes of the Fermi surfaces.
    We note that the branched paths can still be seen when the holes exit and enter the injection and collection QPCs respectively, even though they do not manifest as separate peaks in the conductance spectrum.

    Since the extra peaks are absent when we model the TMF device with QPCs, this lends credence to the hypothesis that those peaks were a product of the abrupt interface mismatch between the semi-infinite leads and the scattering area.
    However, the spin-projected components of the conductance spectra still tell the same story: the peaks are composed of a mixture of $\pm\frac{3}{2}$ and $\pm\frac{1}{2}$ states in either the $J^{}_{z}$ (Fig.~\ref{fig:QPC_TvB}a) or the $J^{}_{x}$ (Fig.~\ref{fig:QPC_TvB}b) basis, with no apparent spin-polarisation.

    We also note that the second peak in the conductance spectrum has a pronounced fringe-like pattern (Fig.~\ref{fig:QPC_TvB}) that was not obvious in Fig.~\ref{fig:TvB_2300nm}.
    This is reflected in the LDOS (Fig.~\ref{fig:QPC_LDOS}b) as well, with interference fringes appearing between the branched paths near the QPCs.
    While fringes in the conductance spectrum can arise due to self-interference between caustic trajectories~\cite{Bladwell2017}, the pattern of those fringes differs significantly from those observed in Fig.~\ref{fig:QPC_TvB}.
    Hence it is unlikely that self-interference is the cause of the fringes in Fig.~\ref{fig:QPC_TvB} and we explore this further in the following sections.

\section{The role of disorder}
\label{section:Disorder}
    If the fringes observed in the conductance spectrum shown in Fig.~\ref{fig:QPC_TvB} were caused by interference effects within the scattering area, they should be affected by disorder.
    We can simulate disorder in our calculations by applying random potentials to each onsite term in the discretised Hamiltonian.
    Here we use a simple Anderson localisation model~\cite{Anderson1958}, where the random onsite potentials are uniformly distributed within a range of $[-U^{}_{0}/2, U^{}_{0}/2]$.
    The mean free path ($l_{\mathrm{mfp}}$) for a hole with a wavevector $k$ traveling through an area of this disorder can be calculated using the equation~\cite{Lee2022}:
    \begin{align}
        l_{\mathrm{mfp}} = \frac{12 \hbar^{4}_{} k}{m^{2}_{\mathrm{eff}} a^{2}_{} U^{2}_{0}}
        \label{Eq:mfp}
    \end{align}
    where $a$ is the discretisation grid spacing, and $m^{}_{\mathrm{eff}}$ is the effective mass.

        \begin{figure}[htp]
        \centering
        \includegraphics[width=0.9\linewidth]{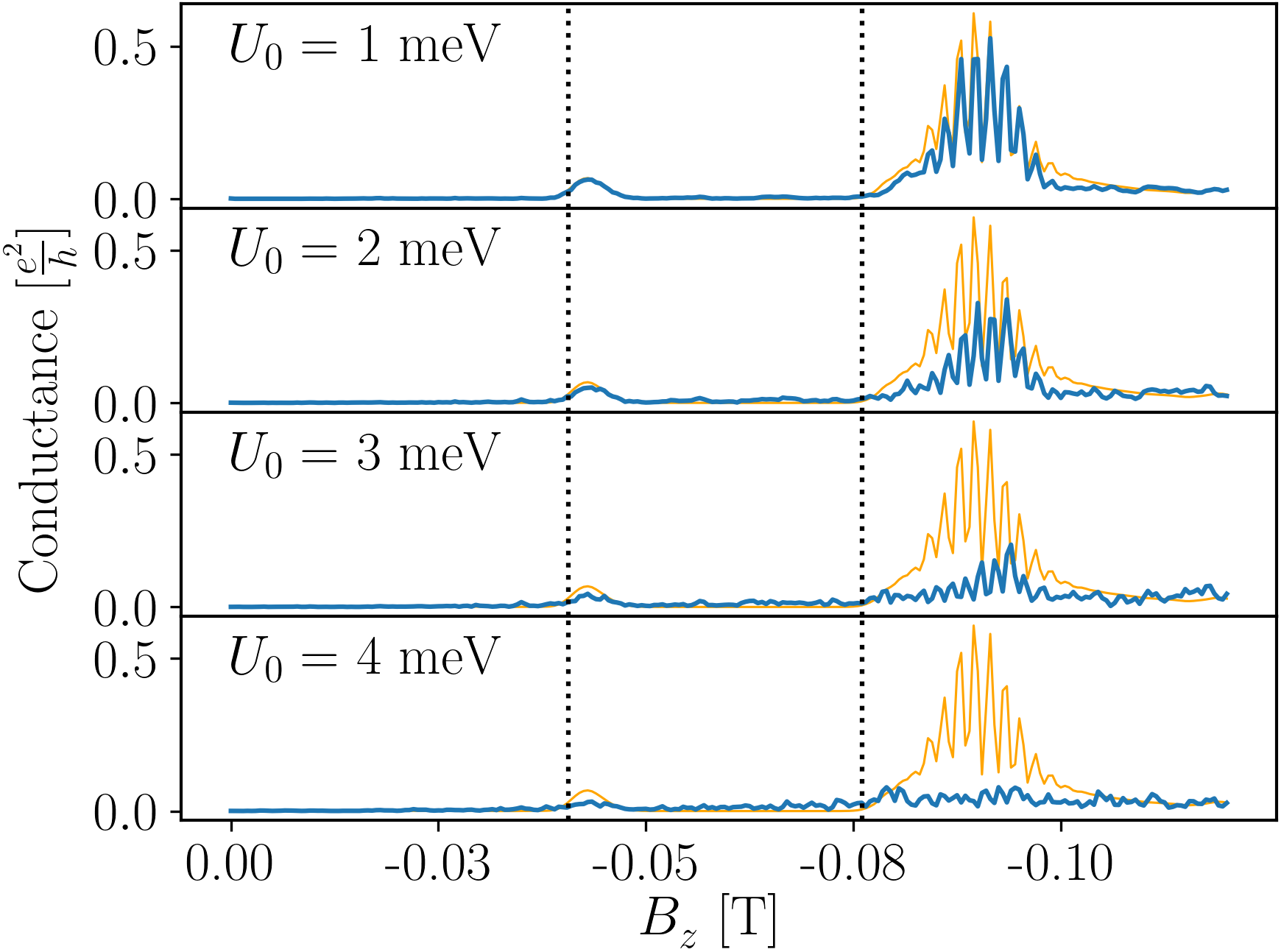}
        \caption{
        Blue solid lines show conductance spectra for a TMF device modelled with QPC leads (Fig.~\ref{fig:QPC_potential}) separated by 2300~nm, calculated with increasing disorder strengths ($U^{}_{0}$).
        The orange line repeated in each plot shows the same conductance spectrum calculated without disorder, taken from Fig.~\ref{fig:QPC_TvB} and shown for comparison.
        The vertical dotted lines indicate the expected focusing magnetic field strengths.
        }
        \label{fig:Disorder_TvB}
    \end{figure}

    Fig.~\ref{fig:Disorder_TvB} shows the conductance spectra calculated using the same device parameters as in Sec.~\ref{section:QPC}, but for each spectrum, a single realisation of disorder was applied with $U^{}_{0}$ set to 1, 2, 3, or 4~meV respectively.
    The total conductance with no disorder from Fig.~\ref{fig:QPC_TvB} is also shown as an orange line in each of the spectra for comparison.
    Since the transport occurs in-plane, we use $m^{}_{0}/(\gamma^{}_{1} - \gamma^{}_{2})$ and $m^{}_{0}/(\gamma^{}_{1} + \gamma^{}_{2})$ as the effective masses that correspond to the inner ($k=$0.07~nm$^{-1}$) and outer ($k=$0.13~nm$^{-1}$) Fermi surfaces respectively, as discussed in Sec.~\ref{subsubsection:removing k_z}.
    When $U^{}_{0} = 1$~meV, the expected mean free path for a hole on the outer Fermi surface (second peak in the conductance spectrum) is approximately 5.89~$\unit{\um}$, comparable to the trajectory length, and the conductance spectrum appears almost identical to the spectrum calculated without disorder.
    When $U^{}_{0} = 2$~meV, the expected mean free path is about 1.47~$\unit{\um}$, which is shorter than the separation distance between the QPCs, and we see that the amplitude of the second peak is reduced by roughly half.
    This attenuation trend of the second peak in Fig.~\ref{fig:Disorder_TvB} continues until $U^{}_{0} = 4$~meV, when $l_{\mathrm{mfp}} \approx 0.37~\unit{\um}$, and the peak is no longer observable.
    The expected mean free path for a hole on the inner Fermi surface (first peak in the conductance spectrum) is 21~$\unit{\um}$ when $U^{}_{0} = 1$~meV and 1.3~$\unit{\um}$ when $U^{}_{0} = 4$~meV, so the first peak is suppressed by disorder at a slower rate compared to the second peak.
    These results and explanation are consistent with the analytic model and experimental data of Rendell et al.~\cite{Rendell2023}, where the authors related the difference in the experimentally measured amplitudes of the focusing peaks for different $D$ to the difference in mass of the two HH subbands.
    
    While this reduction of the conductance with increasing disorder is expected, we note that the fringes in the second peak are still present even with strong disorder, and the positions of the individual fringes match well with the spectrum calculated with no disorder.
    Therefore, it is unlikely that the fringes are a result of interference within the scattering area.

\section{The role of the Rashba contribution}
\label{section:Square well}
    \begin{figure}[htp]
        \centering
        \includegraphics[width=0.9\linewidth]{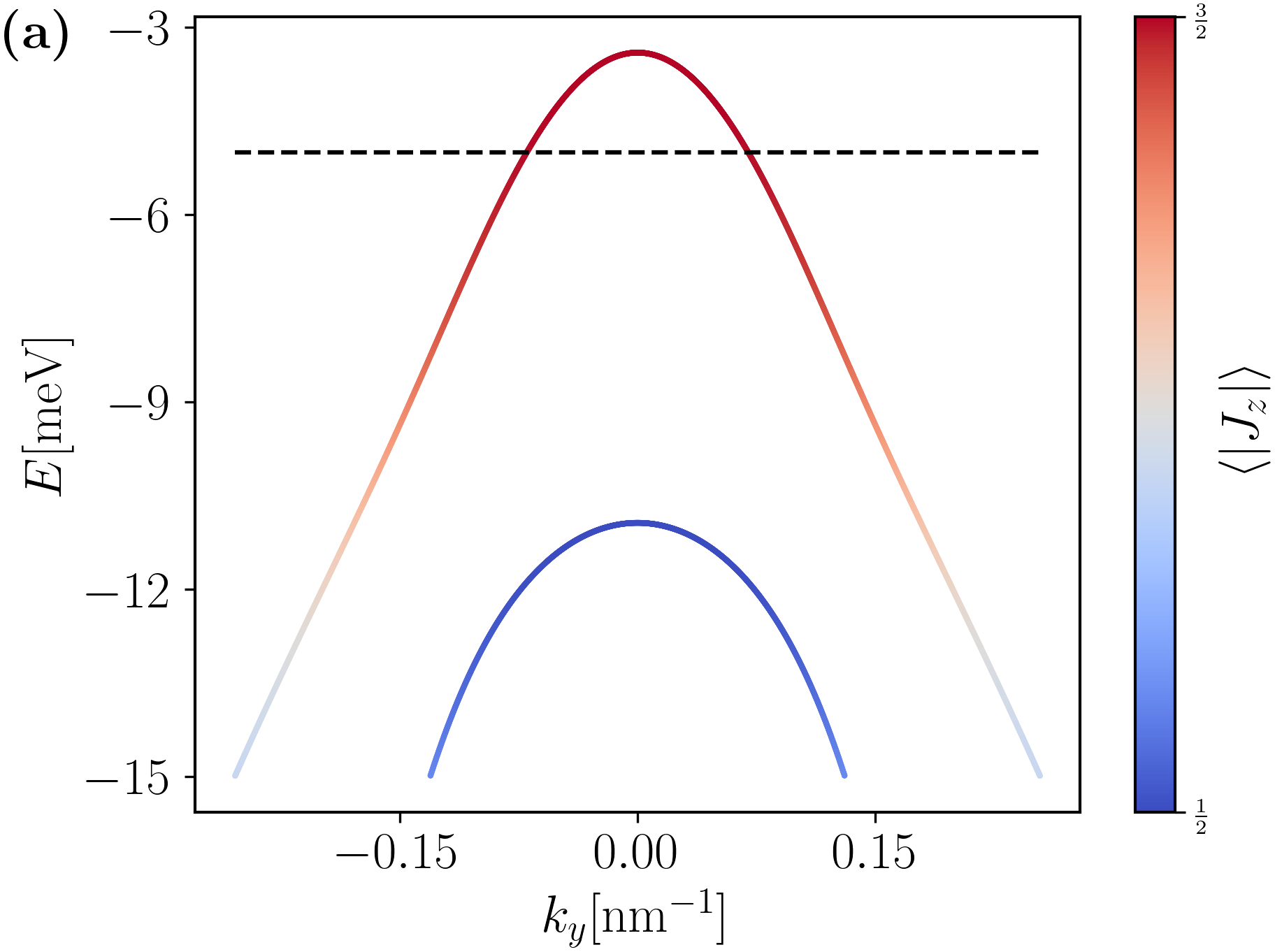}
        \includegraphics[width=0.9\linewidth]{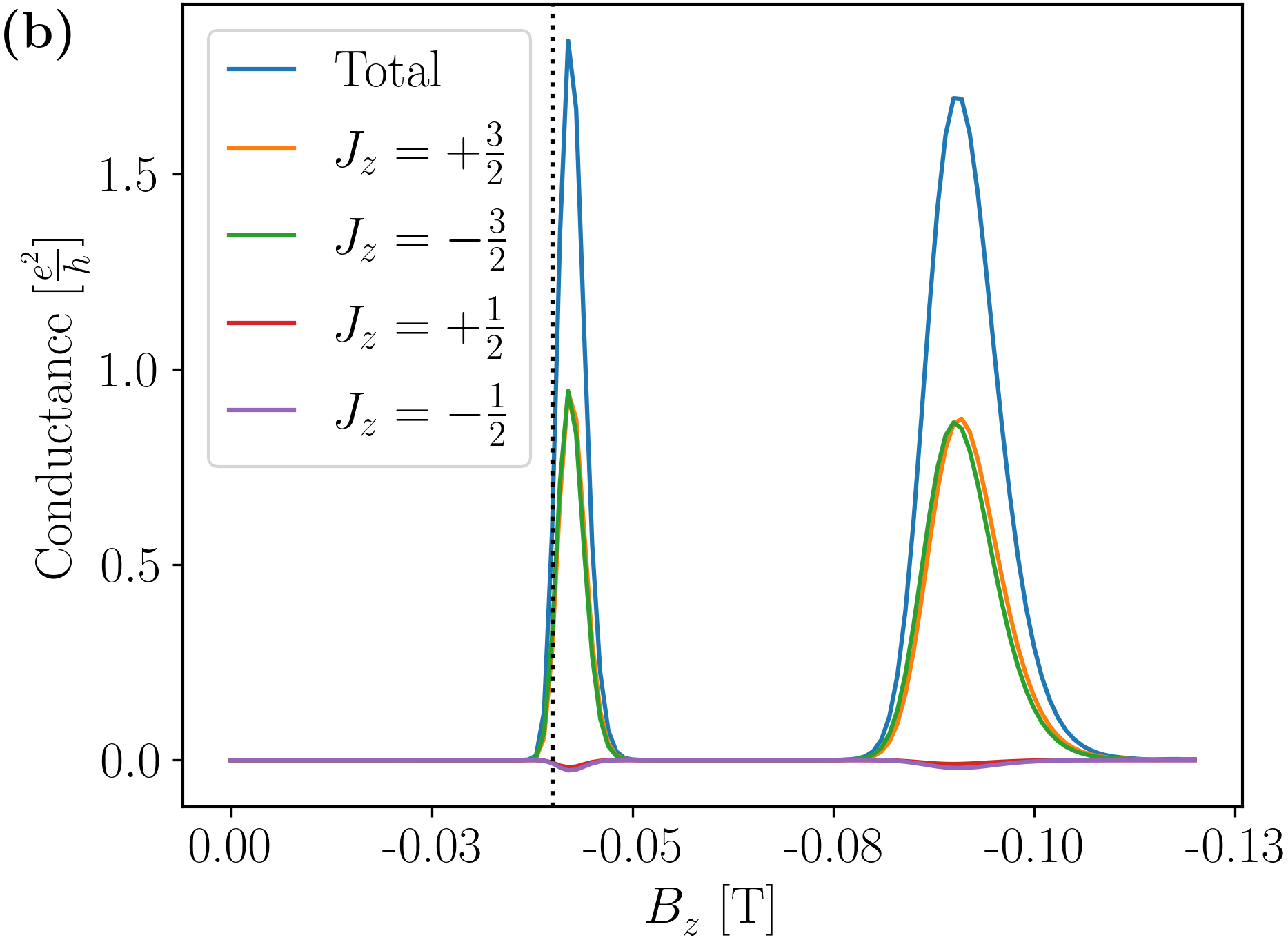}
        \caption{
        (a) Band structure calculated for a GaAs/Al\textsubscript{x}Ga\textsubscript{1-x}As 2DHG confined to the 2D plane by a square well potential, plotted along the $k^{}_{y}$ axis.  
        The colour bar indicates the magnitude of the expectation value of $J^{}_{z}$, where red corresponds to heavy holes ($\langle J^{}_{z} \rangle=\pm\frac{3}{2}$) and blue corresponds to light holes ($\langle J^{}_{z} \rangle=\pm\frac{1}{2}$).
        (b) Conductance spectrum for a TMF device modelled with QPC leads (Fig.~\ref{fig:QPC_potential}) separated by 2300~nm, spin projected in $J^{}_{z}$.
        The confining potential was set to be a square well, and the Fermi energy was set to be approximately $1.6$~meV below the valence band edge, shown by the dashed line in (a).
        The dotted line shows the expected position of the focusing peak according to Eq.~\ref{Eq:larmor}.
        }
        \label{fig:Squarewell_TvB}
    \end{figure}

    An alternative explanation is that the fringes in the conductance spectra are caused by interference occurring within the QPC between the states in the subbands separated by the Rashba effect.
    The Rashba effect is implicitly included in the Hamiltonian (Eq.~\ref{Eq:2DHG_Hamiltonian_2D}) through the asymmetry present in the perpendicular confinement potential (Eq.~\ref{Eq:2DHG_Vz_numerical}).
    Thus we can describe a system without the Rashba effect in our calculations by replacing the perpendicular confinement potential with a square well described by:      
    \begin{align}
        \label{Eq:squarewell_Vz_numerical}
        V(z) =
        \begin{cases}
            \infty & \text{for} \quad z < -\frac{20}{3}W\\
            V^{}_{0} & \text{for} \quad -\frac{20}{3}W < z < 0\\
            0 & \text{for} \quad 0 < z < W\\
            V^{}_{0} & \text{for} \quad W < z < \frac{20}{3}W\\
            \infty & \text{for} \quad z > \frac{20}{3}W
        \end{cases}
    \end{align}
    where we set $V^{}_{0}$ to 211~meV and $W$ to 15~nm, similar to the heterostructure used by Rendell et al.~\cite{Rendell2022}.

    Using Eq.~\ref{Eq:squarewell_Vz_numerical} as the perpendicular confinement potential, we calculated the band structure for a GaAs/Al\textsubscript{x}Ga\textsubscript{1-x}As 2DHG confined to the 2D plane by a square well (as seen in Fig.~\ref{fig:Squarewell_TvB}a).
    Since the HH ($\langle J^{}_{z} \rangle=\pm\frac{3}{2}$) subbands and the LH ($\langle J^{}_{z} \rangle=\pm\frac{1}{2}$) subbands are degenerate respectively, this indicates that there is no Rashba spin-orbit interaction in the system.
    Mixing between the HH and LH subbands also occurs at much higher $k$ values compared to the band structure in Fig.~\ref{fig:Bands_2D}b.

    We then calculated the conductance spectrum using the same device parameters as in Sec.~\ref{section:QPC}, except the perpendicular confinement potential which was replaced with Eq.~\ref{Eq:squarewell_Vz_numerical} and the Fermi energy which was set to approximately 1.6~meV below the valence band edge as indicated by the dashed line in Fig.~\ref{fig:Squarewell_TvB}a.
    We see in the resulting conductance spectrum (shown in Fig.~\ref{fig:Squarewell_TvB}b) that the peaks do not exhibit fringes.
    The spin projections also show that only $\langle J^{}_{z} \rangle=\pm\frac{3}{2}$ states are present, and their degeneracy starts to be lifted at higher magnetic field strengths (the second peak) when the Zeeman effect becomes more significant.
    
\begin{figure}[htp]
        \centering
        \includegraphics[width=0.9\linewidth]{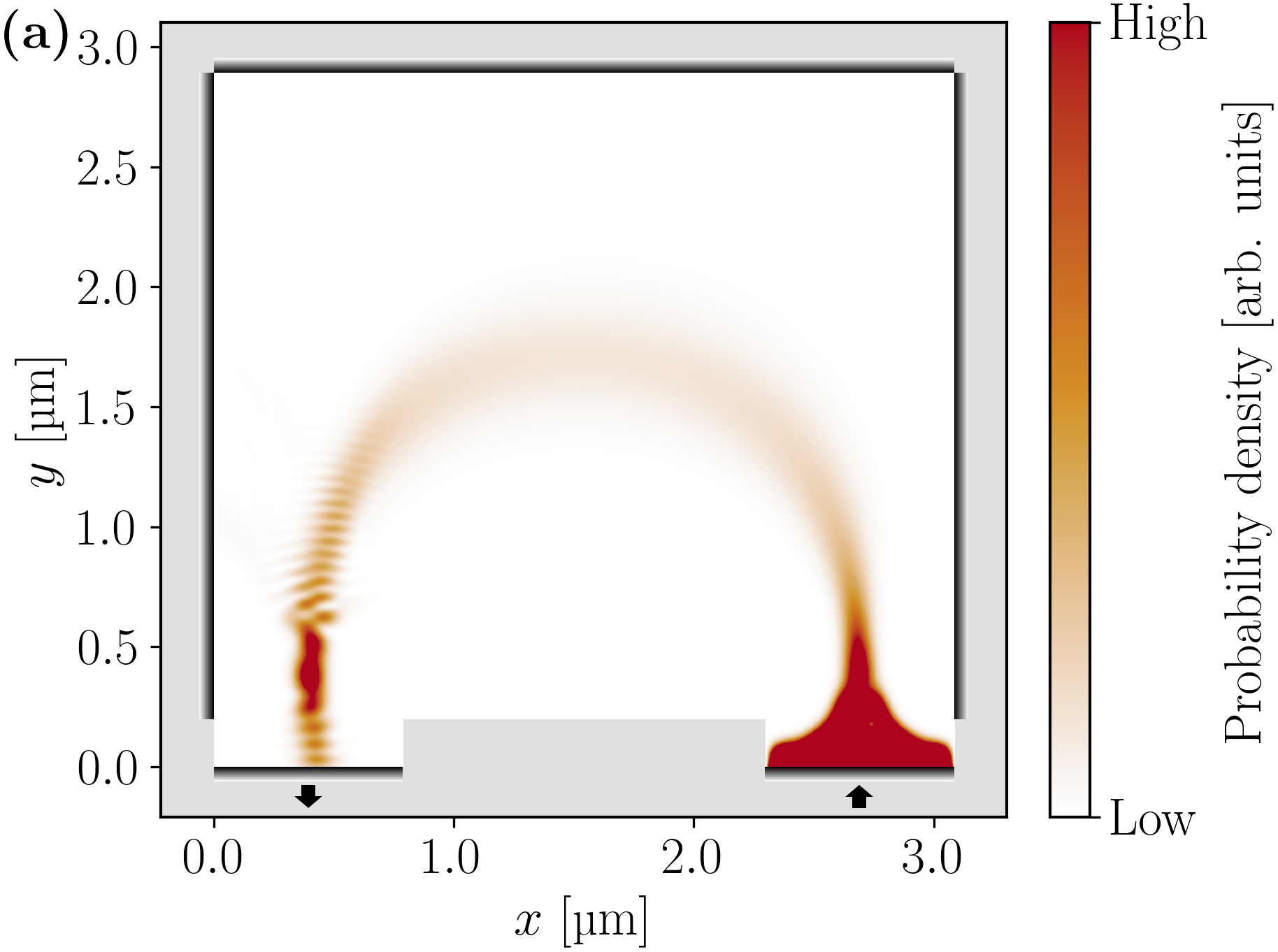}
        \includegraphics[width=0.9\linewidth]{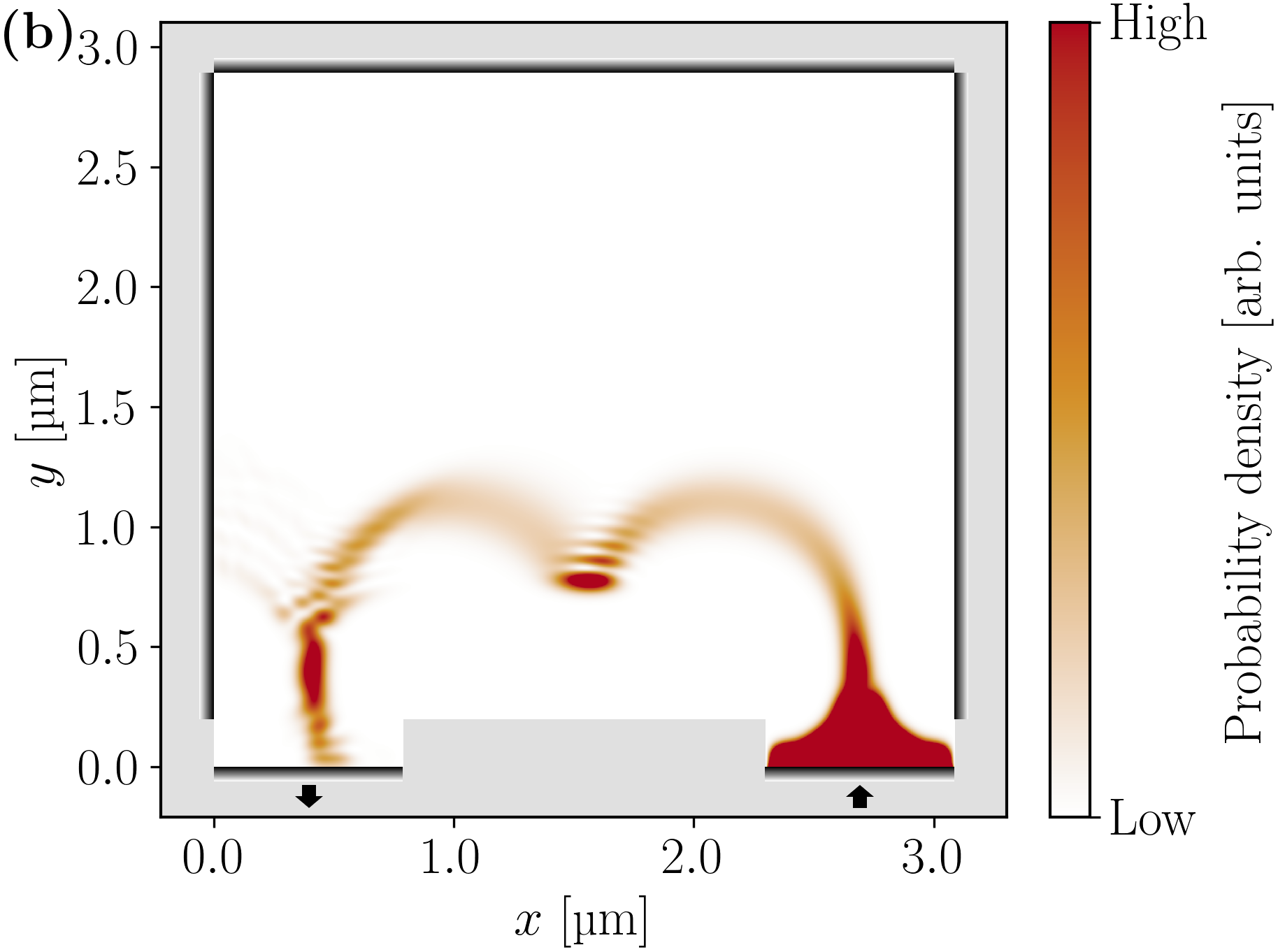}
        \caption{
        LDOS for a TMF device modelled with QPC leads (Fig.~\ref{fig:QPC_potential}) separated by 2300~nm, calculated for (a)~$B^{}_{z} = -0.042$~T and (b)~$B^{}_{z} = -0.091$~T.
        The top gate voltage was set to 10.25~mV, such that the conductance through the injection QPC was one conductance quantum.
        The confining potential was set to be a square well, and the Fermi energy was set to be $1.6$~meV below the valence band edge.
        }
        \label{fig:Squarewell_LDOS}
    \end{figure}

    To further confirm that these two peaks at $B^{}_{z}=-0.042$~T and $-0.091$~T are not caused by a splitting due to the Rashba effect, we also calculated the LDOS at each of those magnetic fields.
    From Fig.~\ref{fig:Squarewell_LDOS}, it can be seen in both plots that the holes only travel in one circular trajectory, instead of separate trajectories that correspond to non-degenerate Fermi surfaces. 
    We can infer that the first and second peaks correspond to when the injected holes have Larmor radii that are approximately one half or one quarter of the separation distance between the QPCs respectively, and thus travel directly from the entry to exit or reflect once off the bottom barrier between the QPCs.
    It is also interesting to note that the branching patterns seen in Fig.~\ref{fig:QPC_LDOS} are not present in Fig.~\ref{fig:Squarewell_LDOS}, further supporting the suggestion that those patterns are caused by the Rashba effect.
    In fact, it appears that in the absence of the Rashba effect, the holes exhibit similar behaviour to electrons.

\section{Summary and outlook}
    Using numerical approaches we modelled a 2DHG TMF device in a GaAs/Al\textsubscript{x}Ga\textsubscript{1-x}As heterostructure with a 4$\times$4 Luttinger Hamiltonian describing the HH-LH subbands.
    By explicitly calculating both the resulting conductance spectra, and the spin resolved charge density, we are able to better understand the spin dynamics and behaviour of the holes within the TMF device. Our results raise a number of pertinent theoretical and experimental questions that will need to be addressed to improve the understanding of transverse magnetic focusing in hole gas devices.

    We found that the ``spin-split'' peaks in the conductance spectra are not actually spin-polarised in the $J^{}_{x}$, $J^{}_{y}$, or $J^{}_{z}$ bases, in contrast to the typical response seen in electron gas experiments. 
    By calculating spin-projected band structures of an infinite 2DHG sheet, we show that this is because mixing between the heavy hole ($\langle J^{}_{z} \rangle = \pm\frac{3}{2}$) and light hole ($\langle J^{}_{z} \rangle = \pm\frac{1}{2}$) states happens at relatively small $k$ in the valence band.
    Our results indicate that accurate interpretation of the measurements from TMF experiments performed in a 2DHG is more complicated that in the corresponding spin-$\frac{1}{2}$ electron gas experiments. 
    Therefore effective Hamiltonians that describe a 2DHG system as an effective spin-$\frac{1}{2}$ system in a pseudospin basis by assuming only the heavy hole states are populated do not fully capture the spin dynamics of the 2DHG system.
    
    Furthermore, the Larmor radius formula as it is written in Eq.~\ref{Eq:larmor} cannot be relied on to accurately predict the position of the peaks in TMF spectra since it assumes a circular trajectory in real space.
    The shape of the outer Fermi surface and our LDOS calculations indicate that holes in a 2DHG can travel in a non-circular trajectory under a perpendicular magnetic field.
    As a result, the effective Larmor radius formula should be generalised to take this behaviour into account.
    To confirm this observation experimentally will require mapping the trajectory of holes under a perpendicular magnetic field, as was done for electrons through scanning probe microscopy~\cite{Aidala2007}.
    
    We also demonstrated that modelling the TMF device with semi-infinite leads directly connected to the scattering area or with QPCs smoothing the transition between the leads and the scattering area gives rise to a stark difference in the results.
    For future numerical work aiming to interpret results of a TMF experiment, it is therefore important to have an accurate characterisation of the QPCs in the experimental device.
    In addition, a semi-analytical analysis of the interference fringes we observed in our calculated conductance spectra (Fig.~\ref{fig:QPC_TvB}) is warranted for a deeper understanding of the behaviour of holes with Rashba spin-orbit interaction in a QPC.
    
    In a similar vein, an accurate description of the perpendicular confinement potential that forms the 2DHG is essential for quantitative numerical models.
    We showed results calculated using both a triangular well, or a square well as the confinement potential - which correspond to a 2DHG with or without Rashba spin-orbit coupling respectively.
    Varying the shape of the perpendicular confinement potential provides a method of exploring the relationship between the behaviour of holes and the Rashba spin-orbit interaction. 
    It would also be feasible to explore this experimentally via both heterostructure engineering and electrical tuning of the Rashba spin-orbit coupling strength~\cite{Rendell2022}.
    
    In conclusion, we have shown that the current interpretations of measurements from 2DHG TMF experiments is insufficient due to the complicated spin dynamics of a spin-$\frac{3}{2}$ system.
    Therefore it is essential to have detailed modelling to accompany experiments so that we can better interpret and understand hole spin dynamics in the future.

\section*{Acknowledgements}
    This work was funded in part by the Australian Research Council Centre of Excellence in Future Low-Energy Electronics Technologies (project number CE170100039).
    Computational resources and services were provided by the National Computational Infrastructure (NCI), which is supported by the Australian Government.
    The authors thank M. Rendell for useful discussions.

\appendix
\section{2DHG band structure projected in individual $J_x$ states}
    \label{Appendix:Jx_bandstructures}
    
    \begin{figure*}[htp]
        \centering
        \includegraphics[width=0.32\linewidth]{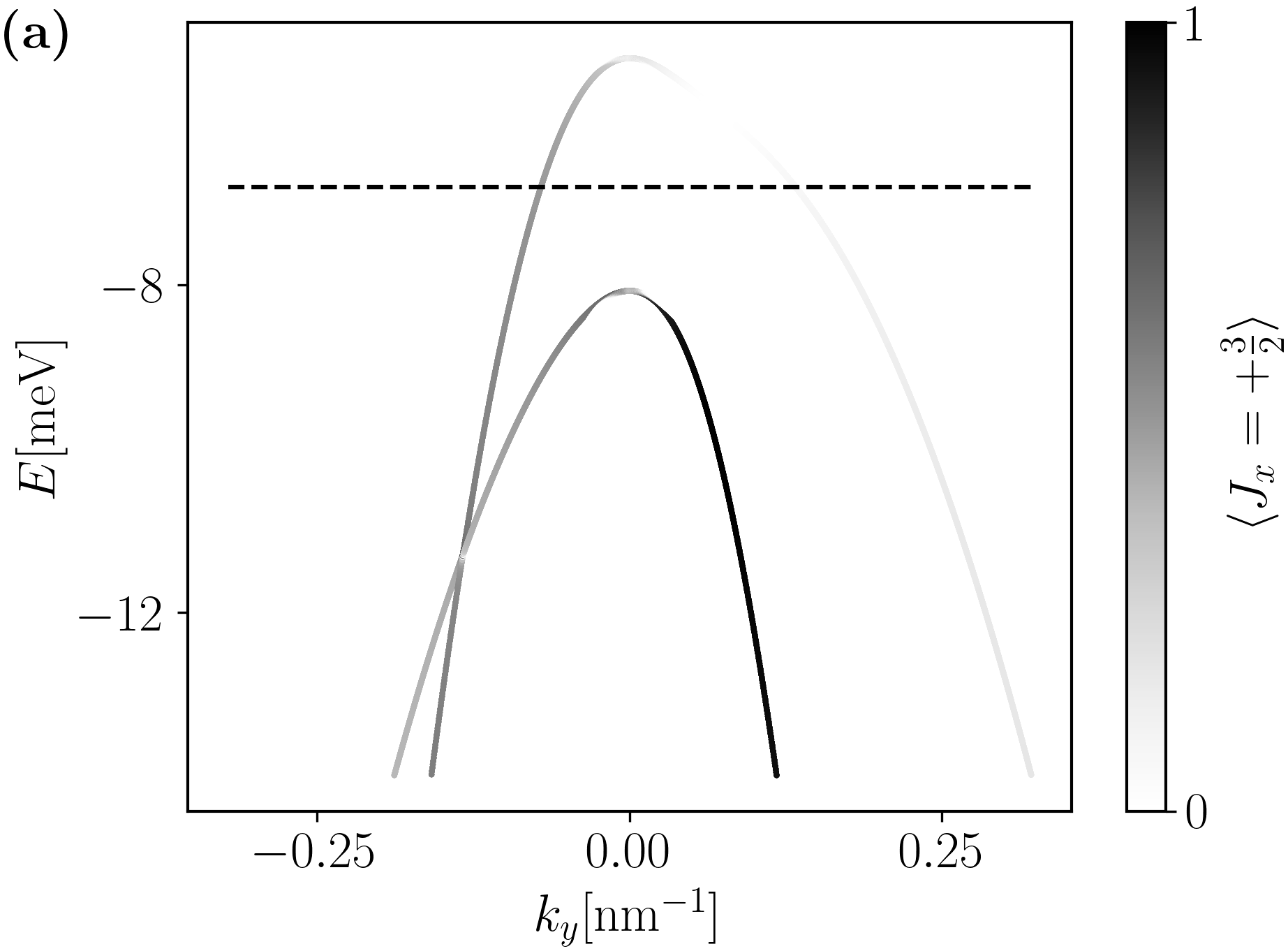}
        \includegraphics[width=0.32\linewidth]{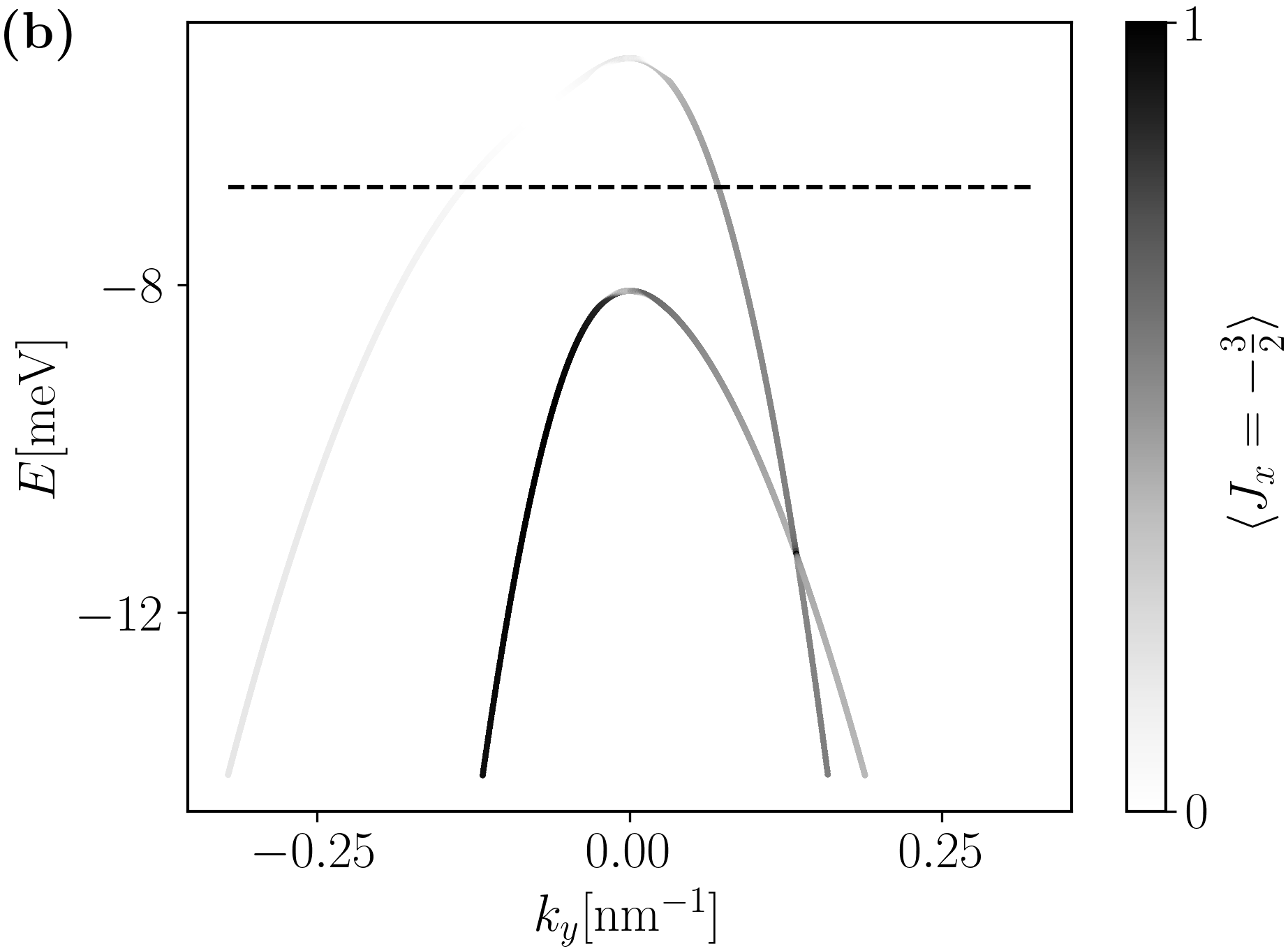}
        \includegraphics[width=0.32\linewidth]{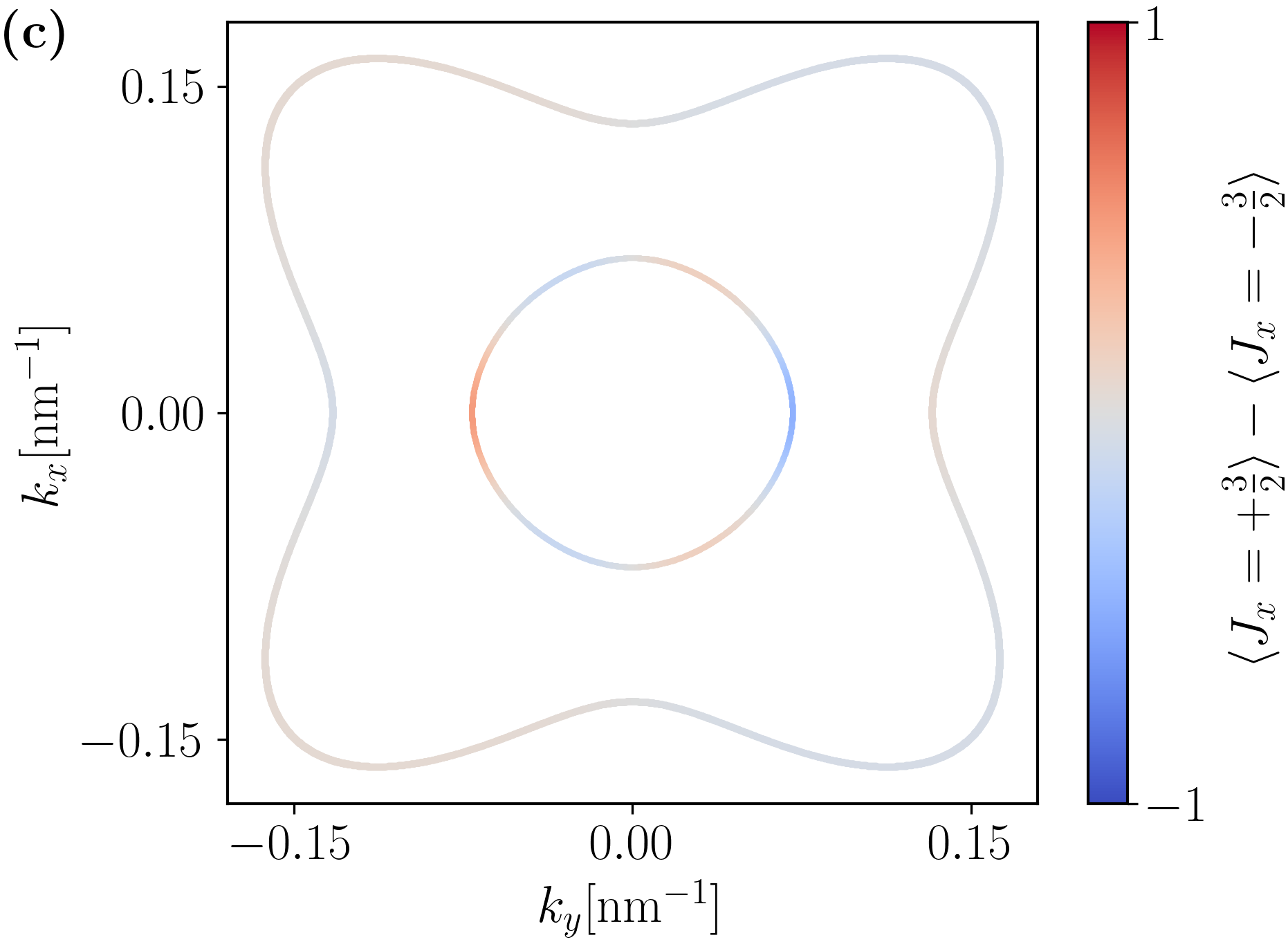}\\
        \includegraphics[width=0.32\linewidth]{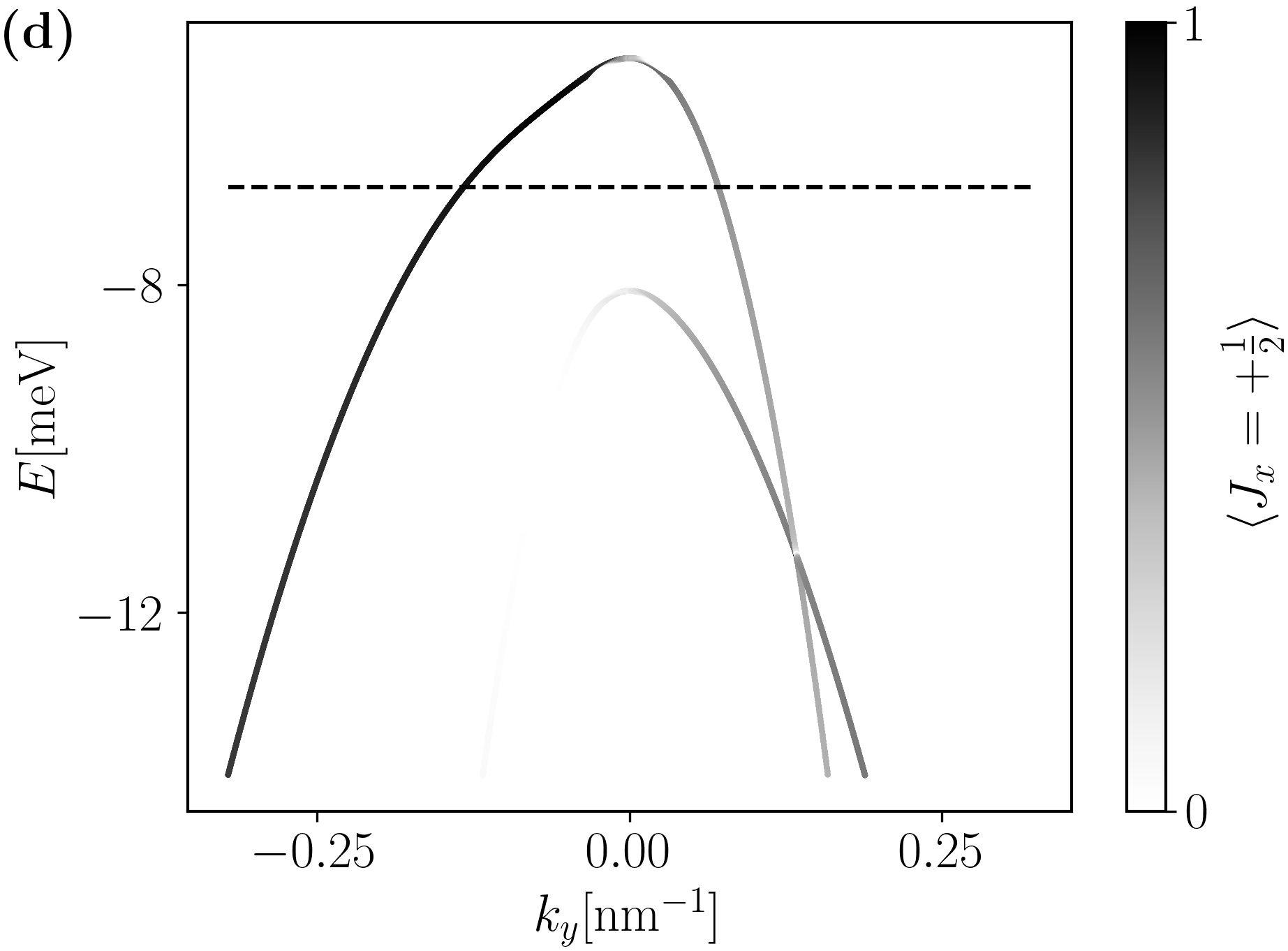}
        \includegraphics[width=0.32\linewidth]{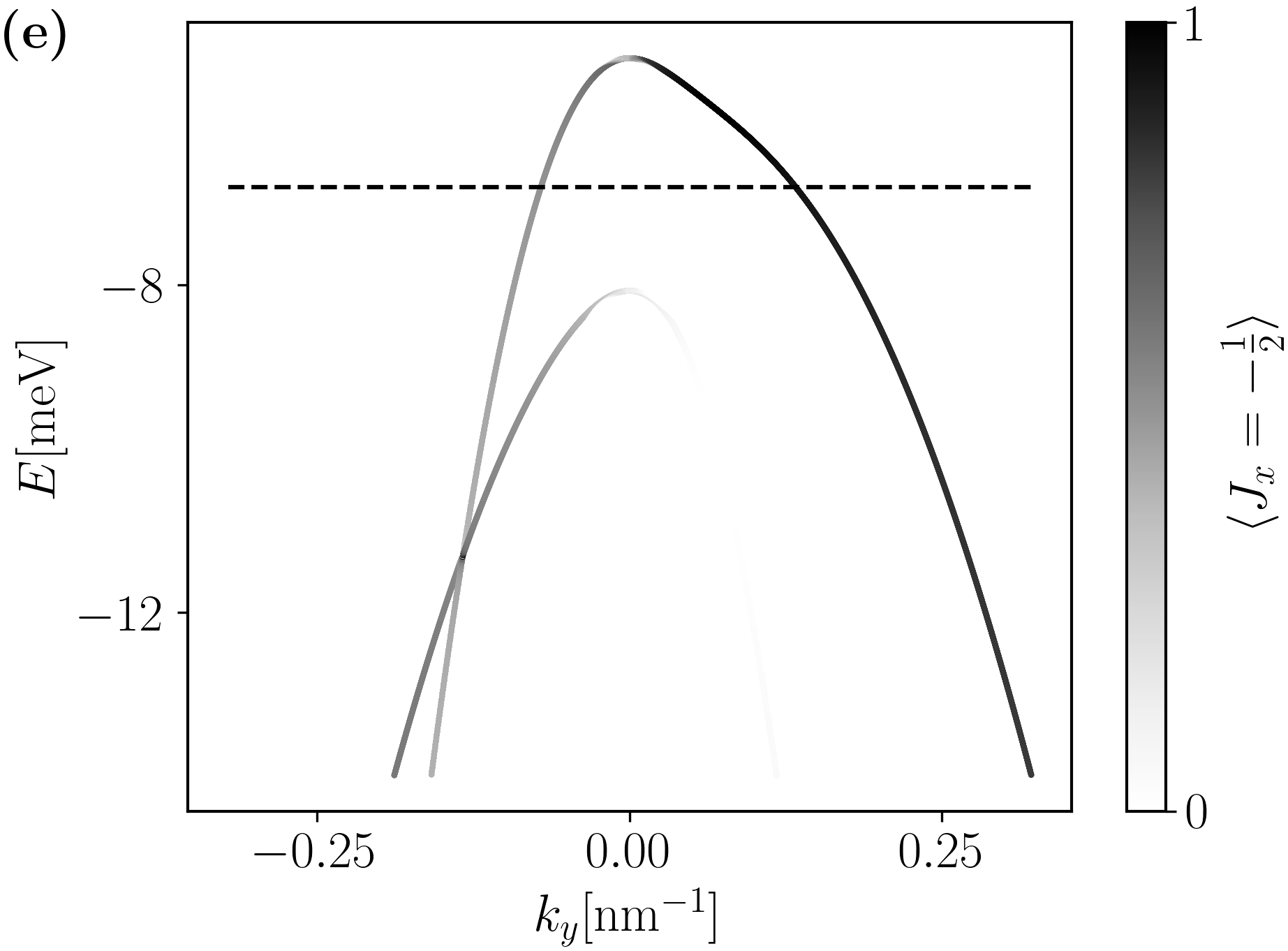}
        \includegraphics[width=0.32\linewidth]{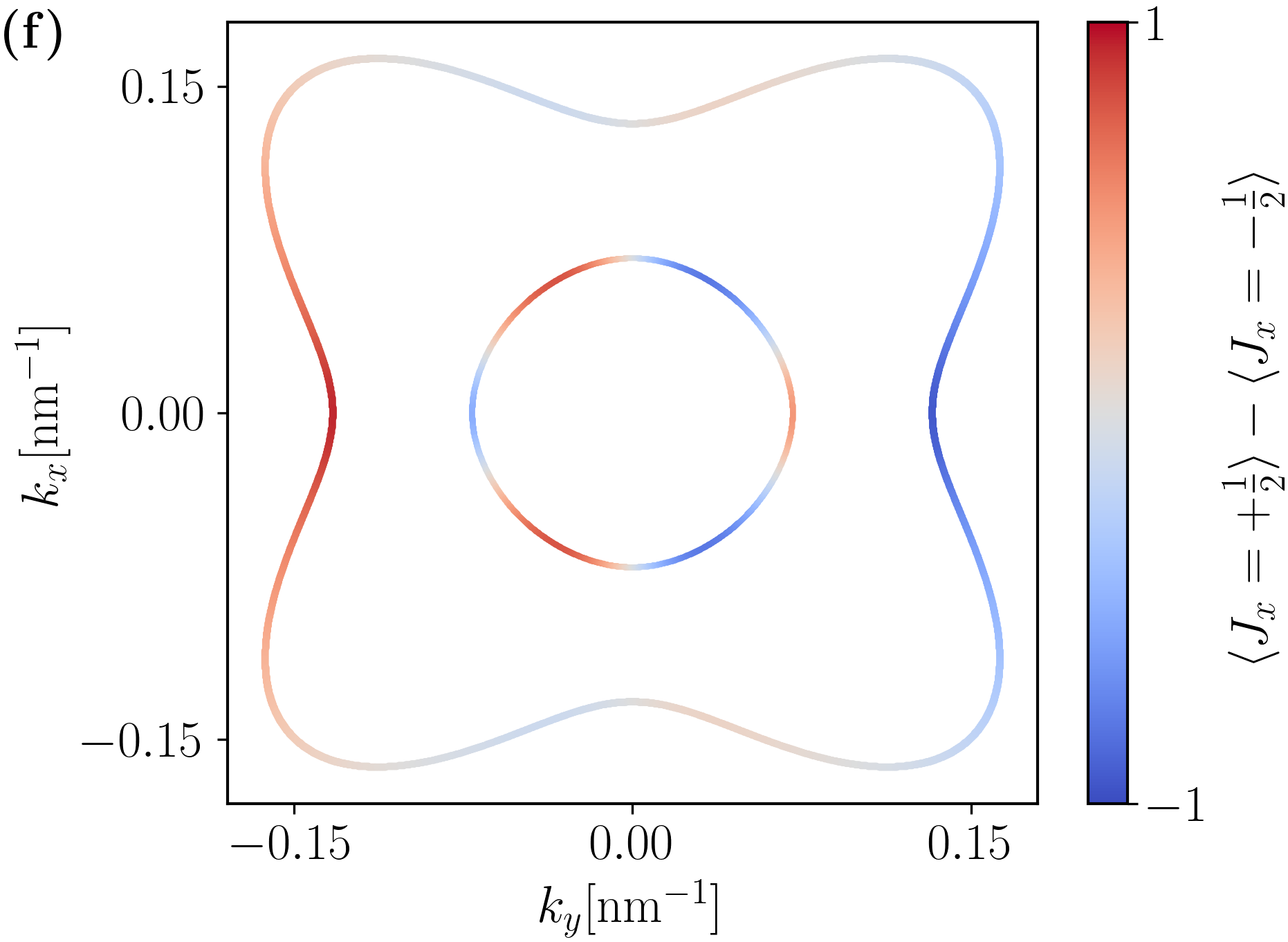}
        \caption{
        Band structures calculated for a 2DHG in GaAs/Al\textsubscript{x}Ga\textsubscript{1-x}As. 
        (a), (b), (d) and (e) are plotted along the $k^{}_{y}$ axis and the dashed lines here mark $1.6$~meV below the valence band edge, the energy at which (c) and (f) were calculated.
        The colour bars in (a), (b), (d) and (e) indicate the probability that $J_x = $ (a) $+\frac{3}{2}$, (b) $-\frac{3}{2}$, (d) $+\frac{1}{2}$, and (e) $-\frac{1}{2}$ at every point in each band structure respectively.
        (c) and (f) show the Fermi surface sliced at $1.6$~meV below the valence band edge.
        The colour bars show the difference between the probabilities that (c) $J_x = +\frac{3}{2}$ and $J_x = -\frac{3}{2}$ and (f) $J_x = +\frac{1}{2}$ and $J_x = -\frac{1}{2}$ respectively, such that red indicates an expected positive spin value and blue indicates an expected negative spin value.
        }
        \label{fig:Jx_bands}
    \end{figure*}

    In Figs.~\ref{fig:Bands_2D}c and~\ref{fig:Bands_2D}f, we saw that the inner and outer subbands projected in the $J^{}_{x}$ basis were not anti-phased. 
    This is in contrast to the anti-phased behaviour seen in a typical 2DEG band structure with Rashba coupling.
    Here we show that this is a consequence of the relative strength of the four $J^{}_{x}$ eigenstates $\left(\pm\frac{3}{2}, \pm\frac{1}{2}\right)$.
    
    Fig.~\ref{fig:Jx_bands} shows the band structure of a 2DHG in a GaAs/Al\textsubscript{x}Ga\textsubscript{1-x}As heterostructure projected in different spin states along the $x$-axis.
    In Figs.~\ref{fig:Jx_bands}a and~\ref{fig:Jx_bands}b, for $k^{}_{y}>0$, the $J^{}_{x} = +\frac{3}{2}$ state is present in the outer HH subband and inner LH subband, while the $J^{}_{x} = -\frac{3}{2}$ state is present in the inner HH subband and outer LH subband.
    This is reversed for $k^{}_{y}<0$, i.e.~the band structure is anti-phased in the $J^{}_{x} = \pm\frac{3}{2}$ projection.
    We can see this more clearly by taking the difference of these two spin-projected band structures at a particular energy, as shown in Fig.~\ref{fig:Jx_bands}c, where $E=1.6$~meV from the valence band edge.
    Although the colours in the outer subband are faint due to the low probability of the $J^{}_{x} = \pm\frac{3}{2}$ states in the outer subband, we can still observe that where the inner subband is blue ($J^{}_{x} = -\frac{3}{2}$), the corresponding section in the outer subband is red ($J^{}_{x} = +\frac{3}{2}$), and vice versa.
    For example, for points on the Fermi surfaces in Fig.~\ref{fig:Jx_bands}c which correspond to $k^{}_{x}=0$ and $k^{}_{y}>0$, the inner band is red and the outer blue, while the reverse is true for $k^{}_{y}<0$.
    Similarly, Figs.~\ref{fig:Jx_bands}d-f show that the inner and outer subbands are anti-phased in the $J^{}_{x} = \pm\frac{1}{2}$ projection as well.

    Note that the probabilities of the $\pm\frac{3}{2}$ states in the HH subbands are lower than the probabilities of the $\pm\frac{1}{2}$ states.
    Also, remember that
    \begin{align*}
        \left\langle J^{}_{x} \right\rangle = &\frac{3}{2} \left\langle J^{}_{x} = +\frac{3}{2} \right\rangle - \frac{3}{2}\left\langle J^{}_{x} = -\frac{3}{2} \right\rangle\\
                                              &+ \frac{1}{2}\left\langle J^{}_{x} = +\frac{1}{2} \right\rangle - \frac{1}{2}\left\langle J^{}_{x} = -\frac{1}{2} \right\rangle
    \end{align*}
    Therefore, we can see how even when the band structure is anti-phased in the $J^{}_{x} = \pm\frac{3}{2}$ and $J^{}_{x} = \pm\frac{1}{2}$ spin projections, the superposition of all the states results in a band structure where the inner and outer subbands not anti-phased as seen in Figs.~\ref{fig:Bands_2D}c and~\ref{fig:Bands_2D}f.

\section{Narrow quantum wells}
    \label{Appendix:narrow_wells}
    
    \begin{figure*}[htp]
        \centering
        \includegraphics[width=0.32\linewidth]{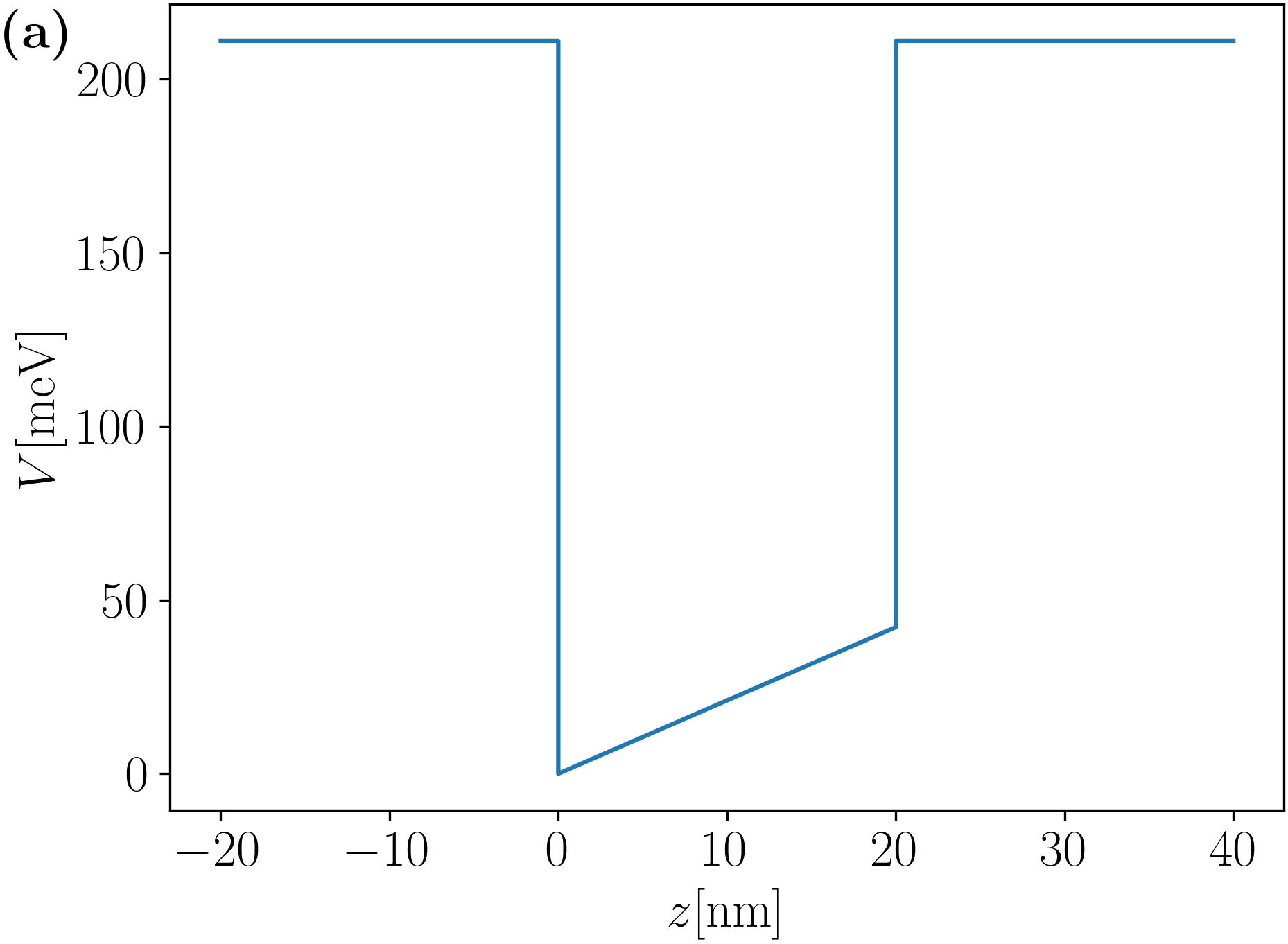}
        \includegraphics[width=0.32\linewidth]{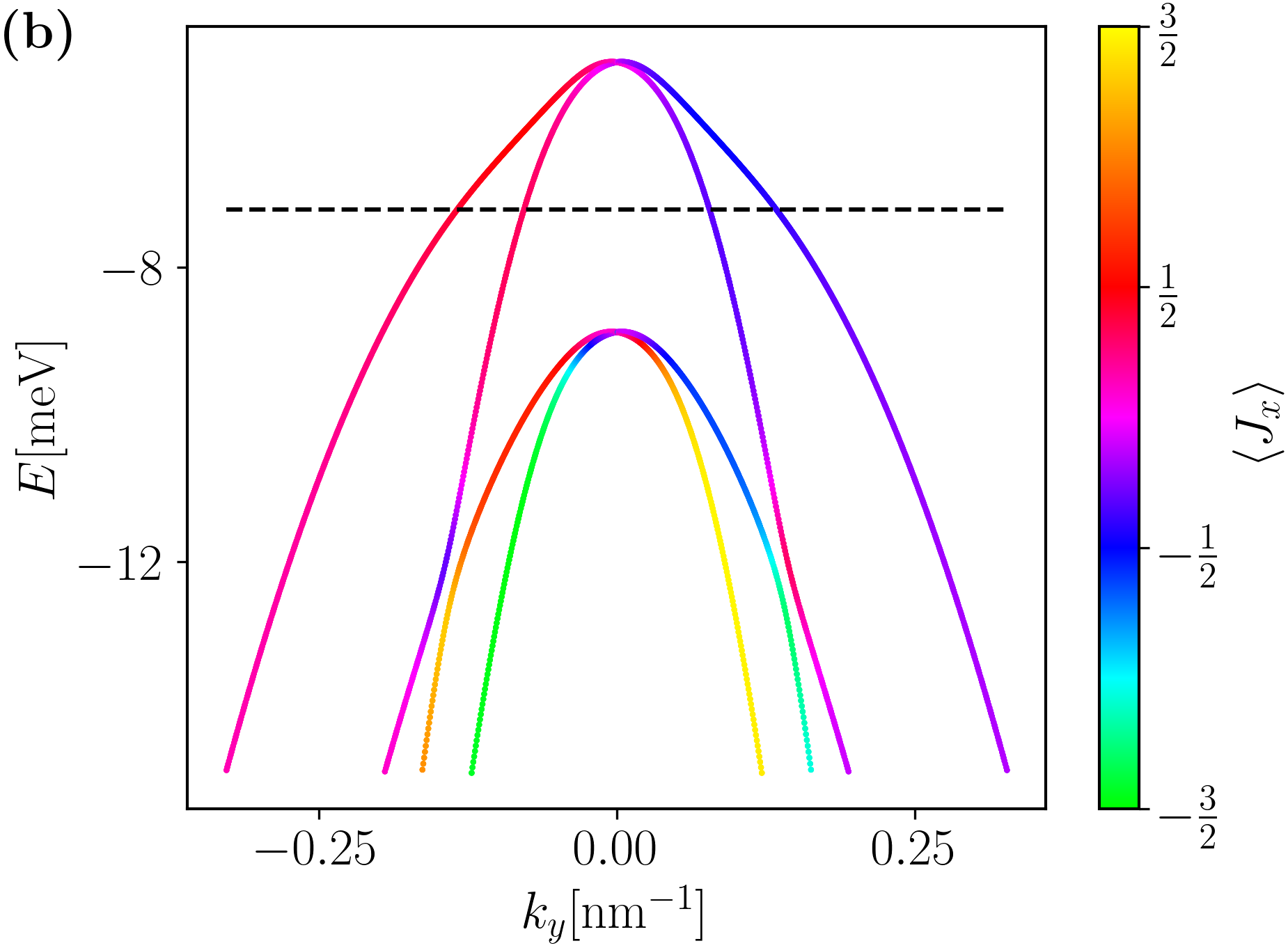}
        \includegraphics[width=0.32\linewidth]{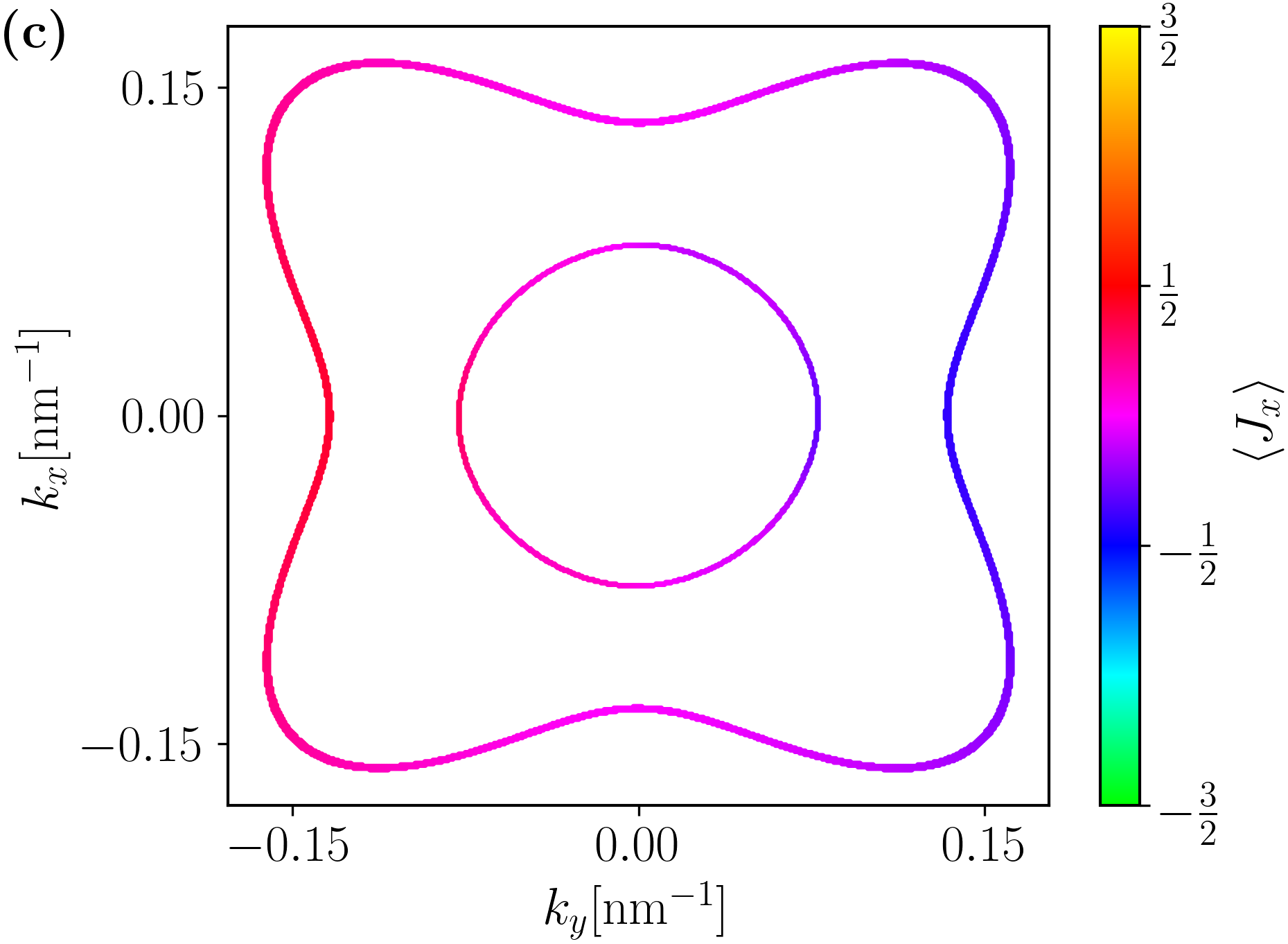}\\
        \includegraphics[width=0.32\linewidth]{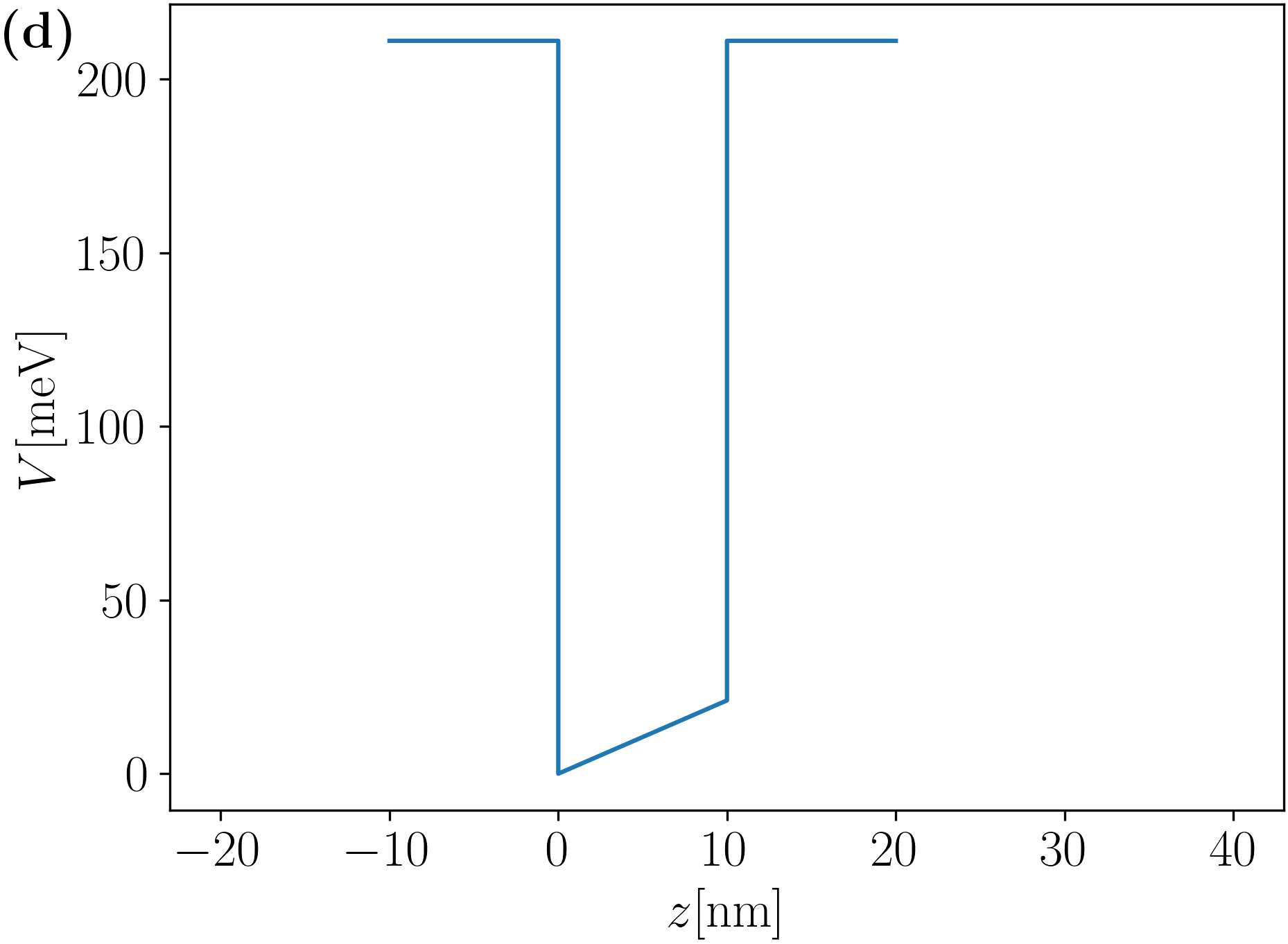}       
        \includegraphics[width=0.32\linewidth]{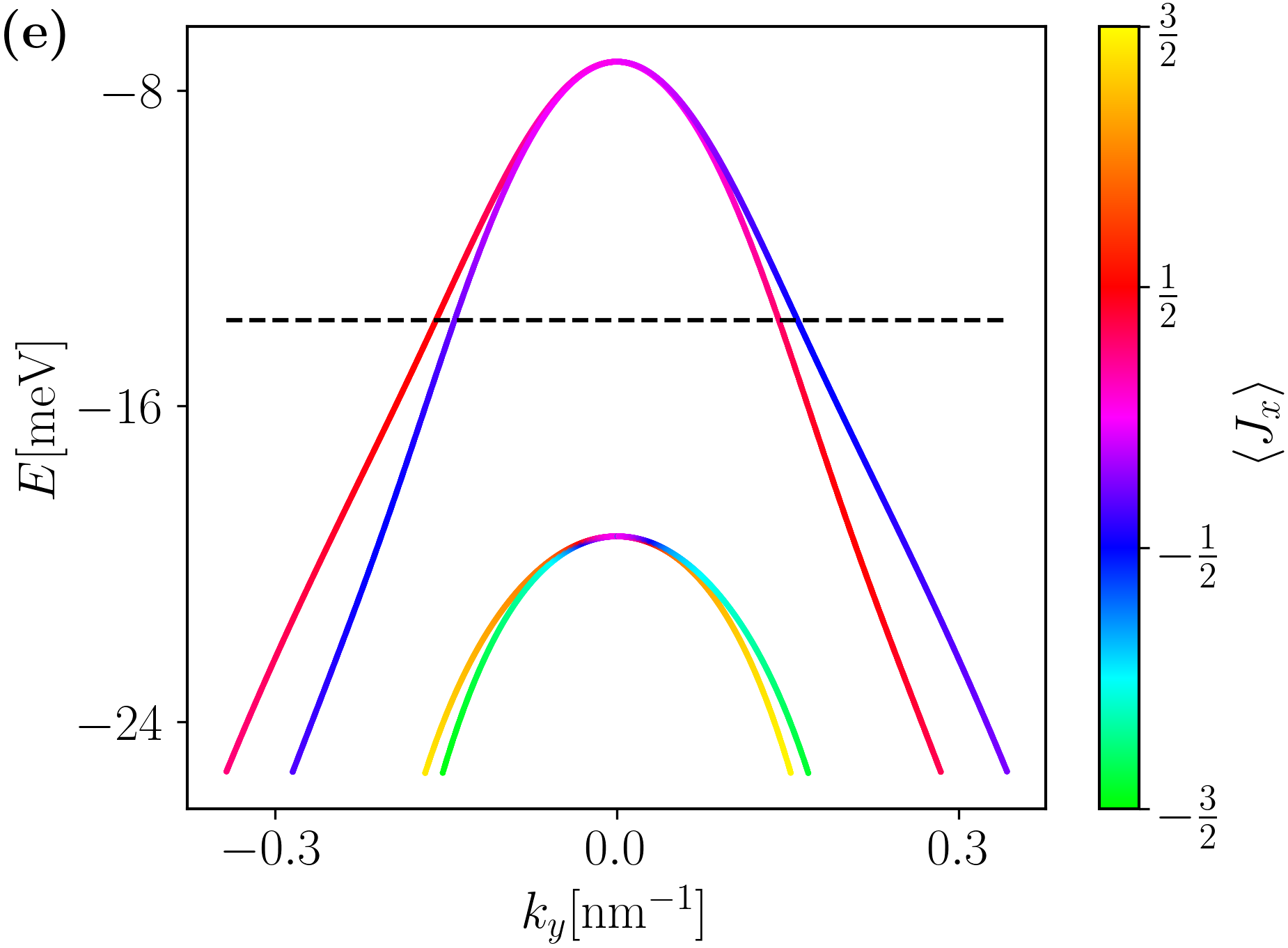}       
        \includegraphics[width=0.32\linewidth]{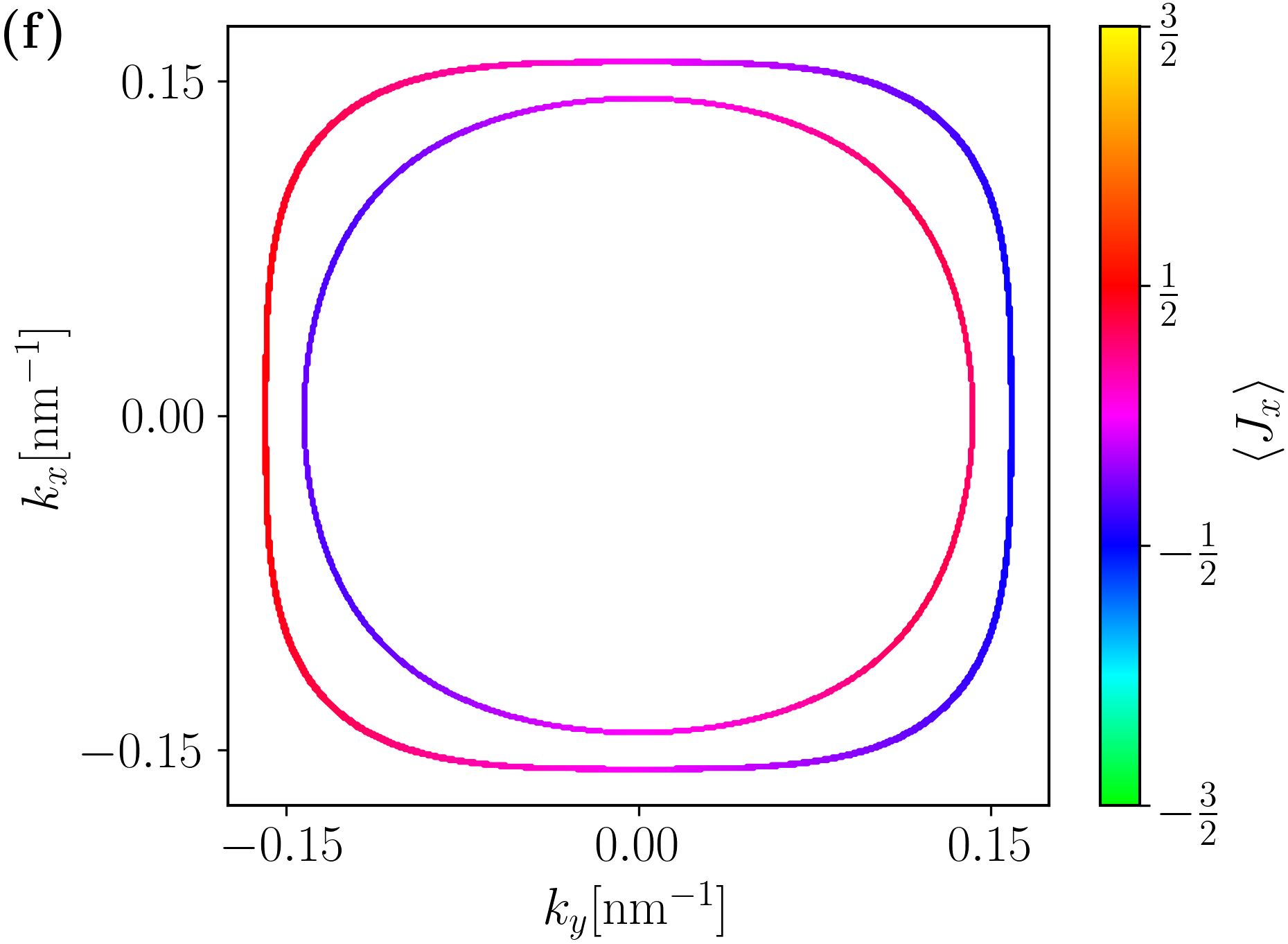}
        \caption{
        (a) Diagram of potential forming a 20~nm wide quantum well that confines a 2DHG, used to calculate the band structures in (b) and (c).
        (d) Diagram of potential forming a 10~nm wide quantum well that confines a 2DHG, used to calculate the band structures in (e) and (f).
        (b) and (e) are plotted along the $k^{}_{y}$ axis, and the dashed lines mark $1.6$~meV below the valence band edge, the energy at which (c) and (f) were calculated respectively.
        (c) and (f) show the Fermi surface sliced at $1.6$~meV below the valence band edge.
        The colour bars show the expectation value of $J^{}_{x}$.
        }
        \label{fig:bands_narrow_wells}
    \end{figure*}
    
    We showed in Sec.~\ref{section:bandstructure} that due to mixing of the HH and LH subbands, the spin-split HH subbands are not spin polarised and thus an effective two-state model would not suffice to describe the spin behaviour.
    Therefore it stands to reason that if the separation between the HH-LH subbands is increased, mixing between the subbands can be avoided and the subbands should become spin polarised again.
    We show in Fig.~\ref{fig:bands_narrow_wells} band structures calculated for 2DHGs confined by quantum wells that are 20~nm and 10~nm wide.
    The shapes of the quantum wells are shown in Figs.~\ref{fig:bands_narrow_wells}a and~\ref{fig:bands_narrow_wells}d.
    These quantum wells can be formed by sandwiching a GaAs layer between two Al\textsubscript{x}Ga\textsubscript{1-x}As layers, as seen in experiments by Rendell et al.~\cite{Rendell2023}, where the GaAs layer was 15~nm thick.
    
    The energy difference between the HH and LH subbands at $k=0$ ($\Delta E_{\mathrm{HL}}$) in Fig.~\ref{fig:Bands_2D}b is approximately 2.96~meV, so the chosen Fermi level was about $0.54\Delta E_{\mathrm{HL}}$ (1.6~meV) below the valence band edge.
    For consistency, we set the Fermi levels in Fig.~\ref{fig:bands_narrow_wells} to be 0.54 times the $\Delta E_{\mathrm{HL}}$ from the valence band edge in each respective band structure.
    We observe in Figs.~\ref{fig:bands_narrow_wells}b~and~c that when the width of the quantum well is 20~nm, the shape of the band structure is still similar to Figs.~\ref{fig:Bands_2D}c~and~f, and the inner and outer subbands are also not anti-phased in spin.
    When the width of the quantum well is reduced to 10~nm, a significant change occurs as we can see in Figs.~\ref{fig:bands_narrow_wells}e~and~f.
    Compared to Fig.~\ref{fig:bands_narrow_wells}c, the shape of the outer subband is less toast-like and more circular, and the inner and outer subbands are now anti-phased in spin.
    Therefore, we would expect that if TMF is performed in a 2DHG formed by a quantum well that is 10~nm wide in $z$, the focusing peaks in the conductance spectra would correspond to spin polarised currents and an effective two-state model can be used to describe the system.






\newpage
\bibliography{2DHG_TMF}


\end{document}